\documentclass[12pt]{article}

\usepackage{graphicx}
\usepackage{float}
\usepackage{cite}
\usepackage{amsfonts}
\usepackage{amssymb}
\usepackage{relsize}
\usepackage[compatibility=false]{caption}
\usepackage{subcaption}

\def\gtwid{\mathrel{\raise.3ex\hbox{$>$\kern-.75em\lower1ex\hbox{$\sim$}}}}
\def\ltwid{\mathrel{\raise.3ex\hbox{$<$\kern-.75em\lower1ex\hbox{$\sim$}}}}
\def\square{\kern1pt\vbox{\hrule height 1.2pt\hbox{\vrule width 1.2pt\hskip 3pt
   \vbox{\vskip 6pt}\hskip 3pt\vrule width 0.6pt}\hrule height 0.6pt}\kern1pt}

\begin{document}

\begin{titlepage}

\begin{flushright}
UFIFT-QG-23-06
\end{flushright}

\vskip 2cm

\begin{center}
{\bf Don't Throw the Baby Out with the Bath Water}
\end{center}

\vskip 1cm

\begin{center}
R. P. Woodard$^{\dagger}$
\end{center}

\begin{center}
\it{Department of Physics, University of Florida,\\
Gainesville, FL 32611, UNITED STATES}
\end{center}

\vspace{1cm}

\begin{center}
ABSTRACT
\end{center}
I stress the importance of retaining a healthy classical limit while 
we search for an ultraviolet completion to quantum gravity. A key problem
with negative-norm quantizations of higher derivative Lagrangians is
that their classical limits do not correspond to real-valued metrics 
evolving in a real-valued spacetime. I also demonstrate that no completion 
based on the flat spacetime background S-matrix can suffice by providing
an explicit example of a theory with unit S-matrix which still shows 
interesting changes in single-particle kinematics and in the evolution 
of its background. I discuss the implications of these considerations 
for the program of Asymptotic Safety. Finally, I urge that some attention 
be given to the possibility that quantum general relativity might make 
sense if only we could go beyond conventional perturbation theory. 

\begin{flushleft}
PACS numbers: 04.50.Kd, 95.35.+d, 98.62.-g
\end{flushleft}

\vskip 2cm

\noindent {\it Dedicated to the memory of Stanley Deser: teacher, mentor, friend.}

\vskip .5cm

\begin{flushleft}
$^{\dagger}$ e-mail: woodard@phys.ufl.edu
\end{flushleft}

\end{titlepage}

\section{A Thought-Provoking Question}

{\it What is wrong with proclaiming that we have solved the problem of quantum
gravity, and the answer is the 1-dimensional simple harmonic oscillator?}
\begin{equation}
H = \frac{p^2}{2 m} + \frac12 m \omega^2 q^2 \; . \label{SHOHam}
\end{equation}
This is an exemplary quantum theory: all its states are normalizable, with 
positive energies and positive norms. Indeed, we can explicitly write down a 
complete set of energy eigenstates with $E_n = (n + \frac12) \hbar \omega$,
\begin{equation}
\psi_n(q) = \Bigl[\frac{m \omega}{\pi \hbar} \Bigr]^{\frac14}
\frac{(a^{\dagger})^n}{\sqrt{n!}}  \exp\Bigl[-\frac{m \omega}{2 \hbar} q^2\Bigr] 
\quad , \quad a^{\dagger} \equiv \sqrt{\frac{m \omega}{2 \hbar}} \, \Bigl( q - 
\frac{\hbar}{m \omega} \frac{d}{d q} \Bigr) \; . \label{SHOstates}
\end{equation}
Of course there is one little problem: our theory of ``quantum gravity'' fails 
to describe the tides --- or the solar system, or gravitational redshift, or the 
precession of the perihelion of Mercury, or the bending of starlight, or Shapiro
delay, or gravitational lensing, or the spin-down of the binary pulsar, or the 
existence of gravitational radiation, or cosmology --- or any of the other 
phenomena which are explained by classical general relativity.

We thereby come to an important realization: {\it any proposal for ``quantum 
gravity'' had better represent the quantization of classical gravitation.} Note 
also that the phenomena I have listed are not described by a gravitational 
S-matrix. All available data \cite{Will:2014kxa} indicate that their explanation 
derives instead from a local, generally coordinate invariant field theory of a 
real-valued metric which exists on a real-valued spacetime, with unobservably 
small quantum gravitational fluctuations at currently attainable energy scales. 
Any proposal for ``quantum gravity'' which fails to obey this Correspondence 
Principle amounts to throwing the baby out with the bath water.

I do not mean to imply that there are no quantum gravitational data. The 
simplest explanation of the primordial scalar power spectrum is as the 
gravitational response to quantum fluctuations in whatever matter field(s) 
drove primordial inflation at a scale as much as 55 orders of magnitude 
higher than the present day \cite{Mukhanov:1981xt}. Something like $10^7$ 
pixels of data on this exist \cite{Planck:2018nkj}, with estimates as high as 
$10^{16}$ potentially recoverable from highly redshifted 21 centimeter
radiation \cite{Loeb:2003ya,Furlanetto:2006jb,Masui:2010cz}. The tensor power
spectrum \cite{Starobinsky:1979ty} is also potentially observable, although
it has yet to be resolved \cite{BICEP:2021xfz}. At the current level of
sensitivity, both spectra can be explained as tree order effects in quantum
general relativity. The full development of 21-cm cosmology may eventually 
permit 1-loop corrections to be resolved \cite{Tan:2021lza}, but these would
arise from the finite, nonlocal part of quantum general relativity regarded
as a low energy effective field theory \cite{Donoghue:1994dn,Donoghue:1995cz,
Burgess:2003jk,Donoghue:2012zc,Donoghue:2017ovt,Donoghue:2022eay}. There is 
no prospect of soon acquiring data which can resolve the ultraviolet 
completion of general relativity.

We should recall why people believe general relativity requires an ultraviolet 
completion. In $D=4$ spacetime dimensions the Lagrangian of the most general, 
invariant, metric theory of dimension four includes four terms,
\begin{equation}
\mathcal{L} = \frac{(R - 2 \Lambda) \sqrt{-g}}{16 \pi G} + \frac{\alpha}{2}
R^2 \sqrt{-g} + \frac{\beta}{2} C^{\rho\sigma\mu\nu} C_{\rho\sigma\mu\nu} 
\sqrt{-g} \; , \label{dim4L}
\end{equation}
where $R$ is the Ricci scalar and $C^{\rho\sigma\mu\nu}$ is the Weyl tensor.
Kelly Stelle long ago proved that a certain quantization of this model is 
perturbatively renormalizable \cite{Stelle:1976gc}, provided all four terms 
are allowed. The first two terms comprise general relativity with a cosmological 
constant, and they not only allowed, but required in order to explain current
data \cite{Will:2014kxa}. The Eddington ($R^2$) term is also permitted as long 
as its coefficient $\alpha$ is positive. Indeed, this term plays a crucial role 
in Starobinsky's model of primordial inflation \cite{Starobinsky:1980te}, which 
so far agrees with all data from the epoch of primordial inflation 
\cite{Planck:2018vyg,Planck:2018jri,BICEP:2021xfz}. The problem is the Weyl 
($C^{\rho\sigma\mu\nu} C_{\rho\sigma\mu\nu}$) term. No matter the sign of 
$\beta$, it gives rise to a massive spin two degree of freedom which renders the 
classical theory virulently unstable \cite{Woodard:2009ns}. {\it The problem of 
quantum gravity is that the Weyl term is not permitted, even though it is 
required for perturbative renormalizability} \cite{DeWitt:1967yk,DeWitt:1967ub,
DeWitt:1967uc,tHooft:1974toh,Deser:1974zzd,Deser:1974cz,Deser:1974cy,
Deser:1974nb,Deser:1974xq,Goroff:1985sz,Goroff:1985th,vandeVen:1991gw}.

The Weyl counterterm would so neatly solve the problem of quantum gravity that
there is a long history of attempts to rehabilitate it \cite{Tomboulis:1977jk,
Salam:1978fd,Antoniadis:1984kd,Antoniadis:1986tu,Johnston:1987ue,Hawking:2001yt,
Salles:2014rua,deOSalles:2018eon,Donoghue:2019fcb,Donoghue:2019ecz,
Mannheim:2020ryw,Donoghue:2021eto,Holdom:2021hlo,Mannheim:2021oat,Mannheim:2023mfp}. 
Many of these efforts involve modifications such as the Lee-Wick mechanism 
\cite{Lee:1969fy,Lee:1970iw,Grinstein:2007mp,Donoghue:2018lmc} and
PT-symmetric quantization \cite{Bender:2007wu,Bender:2008vh}. Section 2 explains
that these violate the Correspondence Principle stated above. Much of the 
pro-Weyl argumentation is based on the assumption that quantum gravity can be 
defined by its perturbative S-matrix on flat space background. Section 3 presents
a nonlinear sigma model whose S-matrix is unity but which still shows interesting 
changes in both its background and in the spectrum of single particle states. 
Section 4 discusses why we really want a theory of quantum gravity, and the 
possibility that it might be nothing more than quantum general relativity. My
conclusions comprise Section 5, which include comments on the consequences of 
the Correspondence Principle for the program of Asymptotic Safety 
\cite{Niedermaier:2006wt,Benedetti:2009rx,Falls:2018ylp}.

\section{Why the Weyl Counterterm Is Forbidden}

There are really three points to make here. The first is that any local, 
invariant, metric extension of general relativity, except for $f(R)$ gravity,
endows the classical theory with a virulent kinetic instability. Second, the 
quantization which was employed in Stelle's proof \cite{Stelle:1976gc} amounts 
to regarding the negative energy creation operators of (\ref{dim4L}) as 
positive energy annihilation operators. The resulting Fock ``states'' have
positive energy but some of them have negative norm. These ``states'' are not 
normalizable under canonical quantization, but might be so in some alternate
quantization scheme. However, and third, the classical limit of the resulting 
theory cannot be any local, invariant, metric theory of gravitation.

\subsection{Why the Classical Theory Is Unstable}

The fundamental problem of higher derivatives has nothing to do with gravity, 
or even field theory, so I will review it in the context of a point particle 
whose position is $q(t)$. Suppose the Lagrangian $L(q,\dot{q},\ddot{q})$ 
contains not just first derivatives $\dot{q}$ but also second derivatives 
$\ddot{q}$. Suppose also that it is {\it nondegenerate}, meaning that 
$\frac{\partial L}{\partial \ddot{q}}$ is monotonic in $\ddot{q}$. 
Ostrogradsky long ago showed that the canonical formalism of such a theory 
requires two coordinates and two momenta \cite{Ostrogradsky:1850fid,
Woodard:2015zca},
\begin{eqnarray}
Q_1 \equiv q \qquad & , & \qquad P_1 \equiv \frac{\partial L}{\dot{q}} -
\frac{d}{dt} \frac{\partial L}{\partial \ddot{q}} \; , \qquad \label{QP1} \\
Q_2 \equiv \dot{q} \qquad & , & \qquad P_2 \equiv \frac{\partial L}{
\partial \ddot{q}} \; . \qquad \label{QP2}
\end{eqnarray}
Note that the assumption of nondegeneracy means we can solve for $\ddot{q}$
as some function $a(Q_1,Q_2,P_2)$, without involving $P_1$. Ostrogradsky's
Hamiltonian is,
\begin{equation}
H\Bigl(Q_1,Q_2,P_1,P_2\Bigr) \equiv Q_2 P_1 + a\Bigl(Q_1,Q_2,P_2\Bigr) P_2
- L\Bigl(Q_1,Q_2,a(Q_1,Q_2,P_2)\Bigr) \; . \label{OstroHam}
\end{equation}
One can easily show that the canonical evolution equations just reproduce
the definitions (\ref{QP1}-\ref{QP2}) and the Euler-Lagrange equation
\cite{Woodard:2015zca}, 
\begin{equation}
\dot{Q}_i = \frac{\partial H}{\partial P_i} \qquad , \qquad 
\dot{P}_i = -\frac{\partial H}{\partial Q_i} \; .
\label{Ostroevol}
\end{equation}

There are a number of things to note about Ostrogradsky's Hamiltonian
(\ref{OstroHam}). First, because it is {\it linear} in the independent canonical 
coordinate $P_1$, $H$ can never be bounded, either below or above. Second,
adding more higher derivatives just makes the problem worse: if the Lagrangian 
contains up to $N$ derivatives of $q$, and is nondegenerate in the highest one, 
then there will be $N$ canonical coordinates $Q_i$ and the Hamiltonian will be 
linear in $(N-1)$ of their conjugate momenta $P_i$. Third, this problem is very 
general, completely independent of any approximation technique such as 
perturbation theory. As long as there are nondegenerate higher derivatives, the 
Hamiltonian cannot be bounded. This is the most compelling explanation for why
Newton's assumption about physics being defined by second order equations of 
motion has not been contradicted in 336 years since the {\it Principia}.

We can now understand why the only permitted extension of general relativity 
is $f(R)$ models. Note first that adding a nondegenerate higher derivative 
makes the Hamiltonian linear in the new conjugate momentum which results. This
means it is unbounded below, but also unbounded above. The usual case is that
the Hamiltonian of the lower derivative theory was bounded below, so the new
higher derivative degree of freedom is negative to make the higher derivative
Hamiltonian unbounded below. But if the lower derivative theory was already 
negative energy then the new, higher derivative degree of freedom would be
positive energy. That is exactly what happens for nonlinear functions of the
Ricci scalar because the only higher time derivative in $R$ is of $h \equiv
\ln[{\rm det}(g_{ij})]$,
\begin{equation}
R = - g^{00} \ddot{h} + f\Bigl(g,\partial g,\partial_i \dot{g}\Bigr) \; .
\label{Rstructure}
\end{equation}
This degree of freedom corresponds to the Newtonian potential, and it would
indeed destabilize general relativity were it not for the Hamiltonian 
constraint. So we get a positive energy degree of freedom when the Lagrangian
depends nonlinearly on $R$. However, unconstrained negative energy degrees of
freedom arise from more general contractions of the Riemann tensor, or from 
derivatives of $R$ \cite{Woodard:2006nt}.

Before closing I should comment on some misconceptions:
\begin{itemize}
\item{The Ostrogradsky problem is a {\it kinetic energy} instability, not a 
potential energy one, so it manifests from the dynamical variable being driven 
towards wild time dependence, not some particular value.}
\item{Negative energy states are no problem unless there are {\it interactions}
which allow balancing positive and negative energy excitations.}
\item{The really crippling instability is driven by the infinite entropy of the
high momentum phase space of an interacting continuum field theory. Indeed, the
only way to get a finite decay rate in this case is by artificially cutting off 
the integration over possible momenta \cite{Cline:2003gs}. On the other hand, 
stable higher derivative theories with a finite number of discrete degrees of 
freedom can exist \cite{Deffayet:2023wdg}.}
\item{The problem is not that the energy decays to negative infinity --- the
energy of any solution is conserved. It is rather that solutions develop a
wild time dependence as more and more positive and balancing negative energy
degrees of freedom are excited. This is why a global constraint on the energy
does no good \cite{Boulware:1983td}. What one needs instead is a constraint on
the local negative degrees of freedom, such as the Hamiltonian constraint which
controls the Newtonian potential.} 
\item{Continuum negative energy particles with a high mass do not decouple from 
low energy physics. Rather they couple more strongly because there are more ways 
to produce them, and balance energy by producing positive energy particles.} 
\item{The Ostrogradskian instability must survive canonical quantization because
it afflicts a large part of the classical phase space.}
\end{itemize}

\subsection{How the Quantum Theory Got Negative Norms}

I have just commented that the Ostrogradskian instability must survive {\it
canonical} quantization. Yet the higher derivative gravity theory in Stelle's 
theorem has no negative energy states \cite{Stelle:1976gc}. Even more surprising,
it does have negative norm states, which of course violates the probabilistic
interpretation of quantum mechanics. How did this happen? The answer is that
Stelle employed a noncanonical quantization, and he was right to do so because
it is only this formulation of (\ref{dim4L}) which leads to a renormalizable
theory.

To understand what happened it is useful to work in the context of a quadratic,
higher derivative oscillator,
\begin{equation}
L = \frac{m}{2 \omega^2} \Bigl[ -g \ddot{q}^2 + \omega^2 \dot{q}^2 - 
\omega^4 q^2\Bigr] \; . \label{quadL}
\end{equation}
The general (classical and quantum) solution is a sum of two oscillators,
\newpage

\begin{equation}
q(t) = \sum_{\lambda = \pm} \Bigl[C_{\lambda} \cos(k_{\lambda} t) + 
S_{\lambda} \sin(k_{\lambda} t) \Bigr] \qquad , \qquad k_{\pm} \equiv \omega 
\Bigl[ \frac{1 \!\mp\! \sqrt{1 \!-\! 4 g}}{2 g}\Bigr]^{\frac12} \; , 
\label{gensol}
\end{equation}
where the canonical coordinate expressions for the coefficients are, 
\begin{equation}
C_{\pm} \equiv \frac{\pm g}{\sqrt{1 \!-\! 4 g}} \Bigl[\frac{k^2_{\mp} Q_1}{\omega^2} 
- \frac{P_2}{g m}\Bigr] \qquad , \qquad S_{\pm} \equiv \frac{\pm g}{\sqrt{1 \!-\! 4 g}} 
\Bigl[ \frac{P_1}{g m k_{\pm}} - \frac{k_{\pm} Q_2}{\omega^2}\Bigr] \; . \label{CSpm}
\end{equation}
The Hamiltonian reveals that the $k_{+}$ mode carries positive energy while the 
$k_{-}$ mode is negative energy, 
\begin{equation}
H = \frac{m}{2} \sqrt{1 \!-\! 4 g} \Bigl[ k_{+}^2 \Bigl(C_{+}^2 \!+\! S_{+}^2\Bigr) 
- k_{-}^2 \Bigl( C_{-}^2 \!+\! S_{-}^2\Bigr) \Bigr] \; . \label{quadH}
\end{equation}
The associated lowering operators are,
\begin{equation}
A_{\pm} = \sqrt{ \frac{m k_{\pm} \sqrt{1 \!-\! 4 g}}{2 \hbar}} \Bigl[ C_{\pm} 
\pm i S_{\pm}\Bigr] \; . \label{pmannihilators}
\end{equation}
Writing the operators in the position representation $Q_i = q_i$ and $P_i = -i\hbar 
\frac{\partial}{\partial q_i}$ allows us to find the normalizable wave 
function\footnote{The normalization factor is,
\begin{eqnarray}
N = \sqrt{\frac{m}{\pi \hbar}} \Bigl[ \sqrt{g} (1 \!-\! 2\sqrt{g})\Bigr]^{\frac14}
\; . \label{normalization} \nonumber
\end{eqnarray}}
they annihilate,
\begin{equation}
\Omega(q_1,q_2) = N \exp\Bigl[ -\frac{m g}{2 \hbar \omega^2} \Bigl( (k_{-} \!-\! k_{+})
k_{+} k_{-} q_1^2 - 2 i k_{+} k_{-} q_1 q_2 + (k_{-} \!-\! k_{+}) q_2^2\Bigr)\Bigr]
\; . \label{truevac}
\end{equation}
Just like any harmonic oscillator (\ref{SHOHam}-\ref{SHOstates}), we can define a 
complete set of energy eigenstates $H \vert N_{+}, N_{-}\rangle = \hbar (N_{+} k_{+}
- N_{-} k_{-}) \vert N_{+}, N_{-} \rangle$,
\begin{equation}
\Bigl\vert N_{+} , N_{-} \Bigr\rangle \equiv \frac{(A_{+}^{\dagger})^{N_{+}} 
(A_{-}^{\dagger})^{N_{-}} }{\sqrt{N_{+} ! \, N_{-} !}} \Bigl\vert \Omega \Bigr\rangle
\qquad \Longrightarrow \qquad \Bigl\langle k , \ell \Bigl\vert k' , \ell'\Bigr\rangle
= \delta_{k k'} \delta_{\ell \ell'} \; . \label{HDGstates}
\end{equation}
This is the {\it canonical} formulation of (\ref{quadL}), and it has negative 
energies (just like the classical theory) and positive norms. It also ``makes sense''
in that the expectation values of positive operators are positive,
\begin{equation}
\Bigl\langle \Omega \Bigl\vert q^2 \Bigr\vert \Omega \Bigr\rangle = 
\frac{\hbar}{2 m \sqrt{g}} \frac{1}{k_{-} \!-\! k_{+}} \qquad , \qquad 
\Bigl\langle \Omega \Bigl\vert \dot{q}^2 \Bigr\vert \Omega \Bigr\rangle =
\frac{\hbar \omega^2}{2 m g} \frac{1}{k_{-} \!-\! k_{+}} \; . \label{norms}
\end{equation}
This theory consists of a real-valued position evolving in real-valued time.
\newpage

Stelle's quantization was based on regarding the negative energy creation operator
$A_{-}^{\dagger}$ as a positive energy annihilation operator. We can easily construct
the state wave function $\overline{\Omega}(q_1,q_2)$ which is annihilated by $A_{+}$
and $A_{-}^{\dagger}$,
\begin{equation}
\overline{\Omega}(q_1,q_2) \propto \exp\Bigl[ -\frac{m g}{2 \hbar \omega^2} \Bigl( 
(k_{+} \!+\! k_{-}) k_{+} k_{-} q_1^2 + 2 i k_{+} k_{-} q_1 q_2 - (k_{+} \!+\! k_{-}) 
q_2^2\Bigr)\Bigr] \; . \label{falsevac}
\end{equation}
I have not given the normalization factor because this wave function is not 
normalizable. It is possible to define a set of energy eigenfunctions $H \vert 
\overline{N_{+} , N_{-}} \rangle = \hbar (N_{+} k_{+} + N_{-} k_{-}) \vert 
\overline{N_{+} , N_{-}} \rangle$ analogous to (\ref{HDGstates}),
\begin{equation}
\Bigl\vert \overline{N_{+} , N_{-}} \Bigr\rangle \equiv \frac{(A_{+}^{\dagger})^{N_{+}} 
(A_{-})^{N_{-}} }{\sqrt{N_{+} ! \, N_{-} !}} \Bigl\vert \overline{\Omega} \Bigr\rangle
\; \Longrightarrow \; \Bigl\langle \overline{k , \ell} \Bigl\vert 
\overline{k' , \ell'} \Bigr\rangle = (-1)^{\ell} \delta_{k k'} \delta_{\ell \ell'} 
\Bigl\langle \overline{\Omega} \Bigl\vert \overline{\Omega} \Bigr\rangle \; . 
\label{falsestates}
\end{equation}
All of these eigenfunctions have positive energy, but the ones with odd $N_{-}$ have
a negative norm. This shows up in the expectation value of the square of the velocity 
being negative,
\begin{equation}
\Bigl\langle \overline{\Omega} \Bigl\vert q^2 \Bigr\vert \overline{\Omega} 
\Bigr\rangle = \frac{\hbar}{2 m \sqrt{g}} \frac{\langle \overline{\Omega} \vert 
\overline{\Omega} \rangle}{k_{-} \!+\! k_{+}} \qquad , \qquad 
\Bigl\langle \overline{\Omega} \Bigl\vert \dot{q}^2 \Bigr\vert \overline{\Omega} 
\Bigr\rangle = -\frac{\hbar \omega^2}{2 m g} \frac{\langle \overline{\Omega} \vert
\overline{\Omega} \rangle}{k_{-} \!+\! k_{+}} \; . \label{falsenorms}
\end{equation}
It is obvious from the negative sign on the velocity that this theory does not 
correspond to a real-valued position moving in real-valued time.

These sorts of eccentricities are standard in noncanonical quantization schemes 
\cite{Mostafazadeh:2003tu,Mostafazadeh:2004mx,Mostafazadeh:2010yx}. Note that they 
have nothing {\it per se} to do with higher derivatives. One could just as easily 
get a purely negative energy set of eigenstates for the simple harmonic oscillator 
(\ref{SHOHam}) by regarding the creation operator as an annihilation operator. Indeed, 
because the Schr\"{o}dinger equation $H \psi(q) = E \psi(q)$ is a second order, 
ordinary differential equation, there are two linearly independent solutions for {\it 
any} energy: positive, negative, imaginary, etc. Normalizability puts the {\it 
quantum} in quantum mechanics.

Stelle is a good physicist and he understood all of this. The reason he quantized
(\ref{dim4L}) the way he did is that perturbative renormalizability requires propagators 
to fall off like the inverse fourth power of the Euclidean momentum. To understand this,
compare the propagator of the canonical theory,
\begin{equation}
i\Delta(t;t') \equiv \Bigl\langle \Omega \Bigl\vert T\Bigl[ q(t) q(t')\Bigr] \Bigr\vert 
\Omega \Bigr\rangle = \frac{\hbar}{m \sqrt{1 \!-\! 4 g}} \Bigl\{ \frac{e^{-i k_{+} 
\vert t - t'\vert}}{2 k_{+}} + \frac{e^{i k_{-} \vert t - t'\vert}}{2 k_{-}} \Bigr\} \; ,
\label{canonprop}
\end{equation}
with the propagator of noncanonically quantized theory,
\begin{equation}
i\overline{\Delta}(t;t') \equiv \Bigl\langle \overline{\Omega} \Bigl\vert T\Bigl[ 
q(t) q(t')\Bigr] \Bigr\vert \overline{\Omega} \Bigr\rangle = 
\frac{\hbar}{m \sqrt{1 \!-\! 4 g}} \Bigl\{ \frac{e^{-i k_{+} \vert t - t'\vert}}{2 k_{+}} 
- \frac{e^{-i k_{-} \vert t - t'\vert}}{2 k_{-}} \Bigr\} \; . \label{falseprop}
\end{equation}
One can express both as integrals over a dummy $0$-component momentum $k_0$ using the
identity,
\begin{equation}
\int_{-\infty}^{\infty} \!\! \frac{d k_0}{2\pi} \frac{i e^{-i k_0 (t - t')}}{k_0^2 \!-\!
k^2 \!+\! i \epsilon} = \frac{e^{-i k \vert t - t'\vert}}{2k} \; . \label{conint}
\end{equation}
The result for the canonical propagator (\ref{canonprop}) is,
\begin{equation}
i\Delta(t;t') = -\frac{\hbar \omega^2}{m g} \int_{-\infty}^{\infty} \!\! \frac{d k_0}{2\pi}
\frac{i e^{-i k_0 (t - t')}}{(k_0^2 \!-\! k_{+}^2 \!+\! i \epsilon) (k_0^2 \!-\! k_{-}^2
\!-\! i \epsilon)} \; . \label{propint}
\end{equation}
The problem is that Euclideanization requires the $k_0$ contour to be Wick rotated into 
the 1st and 3rd quadrants, which will give residues from the factor $(k_0^2 - k_{-}^2 - 
i\epsilon)$,
\begin{eqnarray}
\lefteqn{-\frac{\hbar \omega^2}{m g} \frac{i}{(k_0^2 \!-\! k_{+}^2 \!+\! i \epsilon)
(k_0^2 \!-\! k_{-}^2 \!-\! i \epsilon)} } \nonumber \\
& & \hspace{3.5cm} \longrightarrow -\frac{\hbar \omega^2}{m g} \frac{1}{(k_E^2 \!+\! k_{+}^2) 
(k_E^2 \!+\! k_{-}^2)} + \frac{2\pi \hbar \delta(k_E^2 \!-\! k^2_{-})}{m
\sqrt{1 \!-\! 4 g}} \; . \qquad \label{Euclid}
\end{eqnarray}
In a $3+1$ dimensional field theory the residue terms only fall off like an energy
delta function divided by a single power of the 3-momentum. The advantage of the 
noncanonical propagator (\ref{falseprop}) is that it avoids these residues,
\begin{equation}
i\overline{\Delta}(t;t') = -\frac{\hbar \omega^2}{m g} \int_{-\infty}^{\infty} \!\! 
\frac{d k_0}{2\pi} \frac{i e^{-i k_0 (t - t')}}{(k_0^2 \!-\! k_{+}^2 \!+\! i \epsilon) 
(k_0^2 \!-\! k_{-}^2 \!+\! i \epsilon)} \; . \label{falseint}
\end{equation}

\subsection{Why This Is Bad}

The problem with the negative-norm quantization of higher derivative gravity 
(\ref{dim4L}) is that it violates the Correspondence Principle defined in section 1.
{\it The classical limit of this quantization cannot represent a local, invariant
theory of a real-valued metric existing on a real-valued spacetime.} Before 
discussing how bad this is, please take note of the fact that no amount of further
tinkering with quantum higher derivative theory can avoid it. The fatal minus signs
are {\it required} for the quantum theory to be renormalizable. We also have a 
complete catalog of local, invariant theories of a real-valued metric existing on 
a real-valued spacetime and none of them possess the properties asserted for the 
quantum higher derivative theory.

The test of any physical theory comes in its ability to explain and predict
experiment and observation. I do not rule out the possibility that some alternate
quantization of a classically unstable Lagrangian might do that. For example, the
negative norm quantization of the quadratic model (\ref{quadL}) cannot represent 
the position of a point particle evolving in real time, but that might not pose a
problem if we have no sense of being able to measure such a particle. The problem
for gravity is that, once coordinates are fixed, we do have a strong sense of 
being able to measure a real-valued metric existing on a real-valued spacetime. 
Abandoning canonical quantization for gravity puts everything that we know about 
classical, and even semi-classical, gravitation at risk --- which is {\it 
everything} we know about gravity. It amounts to throwing the baby out with the 
bath water.

I want to stress again that the most stringent test of noncanonical quantization 
is not producing a satisfactory quantum theory. This article began by exhibiting 
an absurd theory of ``quantum gravity'' (\ref{SHOHam}-\ref{SHOstates}) which is 
completely consistent, so there never was any doubt that a satisfactory quantum 
theory could be defined. The real test is whether or not the resulting ``quantum 
gravity'' theory recovers the vast array of phenomena which are explained by 
classical gravitation, including the data on primordial perturbations which 
follow from semi-classical gravitation. Passing this test is automatic for 
canonical quantization but it is highly problematic when alternate quantizations 
are employed. {\it And the burden of proof rests with the people who seek to 
rehabilitate the Weyl counterterm.} Many of the rehabilitators are my friends, 
who have spent as many years as I have in trying to understand quantum gravity. 
I hope they will forgive me if I address them as ``you'' in discussing some of 
the arguments they make that their formulations of quantum gravity have 
satisfactory classical limits. 

I often hear rehabilitators claim that recovering general relativity is 
guaranteed because their low energy perturbative S-matrix is approximately that 
of quantum general relativity. There are a number of problems with this view,
starting with its reliance on the perturbative S-matrix. With the possible
exception of the bending of starlight, known gravitational phenomena are not 
explained using perturbative scattering theory but rather a local field theory 
of a real-valued metric existing on a real-valued spacetime, and often 
responding to nonperturbatively strong sources such as neutron stars. I want to 
see this field theory, and I want to see the {\it exact} formulation of it, not 
what you think is a good approximation. By your own admission, you must add 
something to the S-matrix amplitudes of quantum general relativity to improve 
their behavior at high energies; I want to see the local field theory behind 
the addition, no matter how minor you believe it is.

Another common argument is that the classical theory is that of a real-valued
metric on a real-valued spacetime whose Lagrangian is (\ref{dim4L}). That is
not true. As explained in section 2.1, the classical theory is subject to a
virulent kinetic instability whose lifetime is zero. You claim that your 
alternate quantization produces a theory with only positive energies, and that
this is true for every value of $\hbar$, no matter how small. Hence the 
classical limit of your theory is not that of a real-valued metric on a 
real-valued spacetime whose Lagrangian is (\ref{dim4L}). Please tell me what
it is.

I often hear it asserted that the negative norms only affect very short
wavelength modes, about which we know nothing. This argument ignores the 
distinction between time and space. The negative norm quantization of 
(\ref{dim4L}) results in negative norm modes of high mass, which implies
mode functions with high frequencies, but says nothing about the wave number.
In fact, the negative norm Fourier mode sum includes macroscopic wave lengths,
the same way that the positive norm mode sum does. One consequence is that the
square of the time derivative of the invariant length between spacelike
separated points ($[\int \dot{g}_{ij} dx^i dx^j]^2$) is negative, even for 
macroscopic separations, the same way that $\dot{q}^2$ is negative in 
expression (\ref{falsenorms}). This is not acceptable.

Finally, I have heard it claimed that the classical limit of quantum gravity
is as completely misleading as the classical limit of quantum chromodynamics.
I'm sure the statement was not meant this way, but it sounds like an admission
that my objections to noncanonical quantization schemes for gravity are valid.
I stress again that {\it everything} we know about gravity is either classical 
or, in the case of primordial perturbations, semi-classical. If this is not 
recovered by your theory of ``quantum gravity'' then you have indeed thrown 
the baby out with the bath water. 

\section{Objections to S-Matrix Chauvinism}

Those who seek to rehabilitate the Weyl counterterm often adopt what I call
{\it S-Matrix Chauvinism}, which asserts that the perturbative S-matrix on 
flat space background defines physics and is all we can ever know. Believers
refuse to consider quasi-local fields such as $\int \dot{g}_{ij} dx^i dx^j$,
whose square is negative, even over macroscopic spacelike separations in
the negative norm quantization of (\ref{dim4L}). To them, no problem is real
unless it appears in the S-matrix. Scattering into negative norm states does
not trouble them because these states are unstable, so they cannot really be 
present in the asymptotic scattering space. With some inspired tinkering such 
as the Lee-Wick mechanism, it seems reasonable to them that a unitary and 
causal scattering theory can be defined between positive norm and positive 
energy particles. Hope springs eternal. While not denying the utility of the 
S-matrix, I do dispute the extreme position that local fields have no meaning, 
and also the assertion that the S-matrix suffices to define physics.

\subsection{Abandoning Local Fields Is Not Necessary}

One of my current teaching assignments is graduate electrodynamics. It's a 
fun course on what has to be the greatest story every told in physics: how 
men of genius pieced together mankind's first relativistic, unified field 
theory. I use the classic text by the late J. D. Jackson \cite{Jackson:1998nia}
which takes the reader through a bewildering variety of different solutions
for electric and magnetic fields. No one questions the validity of solving 
for these fields, or what the solutions mean. Yet we are told that it all
becomes meaningless nonsense as soon as quantum effects are turned on, no 
matter how small they are. For example, quantum gravity makes a fractional 
correction of about $10^{-45}$ to the Coulomb potential of an electron at
the Compton radius. The best human technology cannot measure an effect so 
small, yet we are told that it being anything other than zero means we must
abandon local fields and base physics entirely on the S-matrix. That never 
made any sense to me.

The alleged incompatibility of quantum effects and local fields is certainly
not for lack of a quantum generalization of the classical Maxwell equation. 
The required generalization is based on the evocatively named ``vacuum 
polarization'' $i[\mbox{}^{\mu} \Pi^{\nu}](x;x')$, which is the 1PI 2-point 
function of the photon,
\begin{equation}
\partial_{\nu} F^{\nu\mu}(x) + \int d^4x' \Bigl[\mbox{}^{\mu} \Pi^{\nu}
\Bigr](x;x') A_{\nu}(x') = J^{\mu}(x) \; . \label{QMax}
\end{equation}
Here $F_{\mu\nu} \equiv \partial_{\mu} A_{\nu} - \partial_{\nu} A_{\mu}$ is
the field strength tensor and $J^{\mu}(x)$ is the current density. S-matrix
chauvinists disparage this equation for three reasons:
\begin{itemize}
\item{$[\mbox{}^{\mu} \Pi^{\nu}](x;x')$ is not real;}
\item{$[\mbox{}^{\mu} \Pi^{\nu}](x;x')$ is nonzero for $x^{\prime \mu}$ outside 
the past light-cone of $x^{\mu}$; and}
\item{The $[\mbox{}^{\mu} \Pi^{\nu}](x;x')$ induced by quantum gravity is 
highly gauge dependent.}
\end{itemize}
All three of these problems occur in the dimensionally regulated, primitive
contribution from a single loop of gravitons on flat space background
\cite{Leonard:2012fs},
\begin{equation}
i\Bigl[\mbox{}^{\mu} \Pi^{\nu}\Bigr](x;x') = -\frac{\kappa^2 \mathcal{C}_0(D,a,b)
(D\!-\!2) \Gamma^2(\frac{D}2 \!-\! 1)}{32 (D\!-\! 1) \pi^D} \Bigl[\eta^{\mu\nu}
\partial^2 - \partial^{\mu} \partial^{\nu}\Bigr] \frac1{\Delta x^{2D-2}} \; .
\label{primPi}
\end{equation}
Here $\kappa^2 \equiv 8 \pi \hbar G/c^3$ is the loop-counting parameter of 
quantum gravity, the Lorentz interval is,
\begin{equation}
\Delta x^2 \equiv \Bigl\Vert \vec{x} - \vec{x}' \Bigr\Vert^2 - \Bigl( 
c \vert t - t'\vert - i\epsilon\Bigr)^2 \; , \label{Dx2}
\end{equation}
and the dependence on the two covariant gauge parameters $a$ and $b$ is,
\begin{equation}
\mathcal{C}_0(a,b,D) = 8 + \frac{12 (a\!-\!1) - 24 (b \!-\!1)^2}{(b\!-\!2)^2} 
+ O(D\!-\!4) \; . \label{Cdef}
\end{equation}

The first two problems derive from the vacuum polarization of expression 
(\ref{primPi}) being an in-out matrix element appropriate to asymptotic 
scattering theory. Using it in the effective field equation (\ref{QMax})
gives the in-out matrix element of $F_{\mu\nu}$, which is not necessarily 
real, even if the field operator is Hermitian, because the in and out vacua 
might differ. In 1960 Julian Schwinger devised a diagrammatic procedure 
\cite{Schwinger:1960qe,Mahanthappa:1962ex,Bakshi:1962dv,Bakshi:1963bn,
Keldysh:1964ud,Chou:1984es} for computing true expectation values which 
is almost as easy to use as the Feynman rules are for computing in-out
matrix elements. When the associated vacuum polarization is employed in
equation (\ref{QMax}) the resulting solutions are real, and the only
contributions from the integration over $x^{\prime\mu}$ are on or inside
the past light-cone of $x^{\mu}$ \cite{Jordan:1986ug,Calzetta:1986ey}.
Further, it is very simple to convert the in-out vacuum polarization to
the in-in one \cite{Ford:2004wc}.

That leaves only the gauge problem. One can see from expression (\ref{Cdef})
that by taking the parameter $b$ close to two and varying the parameter $a$,
we can make the quantum correction run all the way from minus infinity to
plus infinity. This is clearly unacceptable. It is also the result of a 
mistake. The effective field must be excited by some physical source, and
it must be detected by some physical observer. The source and observer 
interact with gravity because all things do. Ignoring the resulting quantum
gravitational correlations is what causes the gauge dependence of the 
effective field equation \cite{Miao:2017feh}. When proper account is taken
of these correlations the renormalized result is a completely gauge independent 
effective field equation which is manifestly real and causal 
\cite{Katuwal:2020rkv},
\begin{eqnarray}
\lefteqn{\partial_{\nu} F^{\nu\mu}(x) + \frac{5 \hbar G \partial^6}{48 \pi^2 c^3} 
} \nonumber \\
& & \hspace{0cm} \times \!\!\int \!\! d^4x' \theta(c \Delta t \!-\! \Delta r) 
\Bigl\{\ln\Bigl[ \mu^2 \Bigl(c^2 \Delta t^2 \!-\! \Delta r^2\Bigr) \Bigr] - 1 
\Bigr\} \partial'_{\nu} F^{\nu\mu}(x') = J^{\mu}(x) \; . \qquad \label{finalQMax}
\end{eqnarray}
Here $\Delta t \equiv t - t'$ and $\Delta r \equiv \Vert \vec{x} - \vec{x}'\Vert$.
This equation can be solved the same way the classical Maxwell equation is, and
resulting electric and magnetic fields have the same transparent physical 
interpretations as in classical electrodynamics. It was never necessary to 
base physics on the S-matrix.

\subsection{Doubts about the S-Matrix}

All of which raises the question of whether or not it is even possible to base
physics on the S-matrix. I have always thought this dubious for a theory of long
range interactions such as gravity. Indeed, infrared divergences preclude the
existence of a gravitational S-matrix on flat space background, although inclusive 
rates and cross sections do exist \cite{Weinberg:1965nx}. Not all theories even
have these. This is especially true for systems in which the background continues
evolving at late times, such as a scalar with a cubic self-interaction 
\cite{Veneziano:1972rs}. S-matrix chauvinists dismiss vacuum decay, but I refuse 
to accept that there are no interesting quantum field theory questions to pose 
about such systems. The persistence of evolution for all time sounds a lot like 
what happens in cosmology.

Cosmology poses insurmountable problems for S-matrix chauvinists because the
universe did not begin in free vacuum but rather with an initial singularity,
and the phenomenon of cosmological particle production means that it cannot 
end in free vacuum. A formal S-matrix can be defined for massive fields on de 
Sitter background \cite{Marolf:2012kh}, but causality renders it unobservable.
And the construction altogether fails when applied to realistic geometries, 
or when one attempts to include massless fields which are not conformally 
invariant such as the graviton.

Cosmology also poses problems for the claim that 
massive, negative norm particles disappear from physics because they are
unstable. The massive scalar degree of freedom in Starobinsky's $R + R^2$ 
model of inflation \cite{Starobinsky:1980te} is also unstable so, by the 
logic of the S-matrix chauvinists, it too must be absent from the space 
of scattering states. Yet the initial value data associated with this 
degree of freedom matters --- it controls the duration of primordial 
inflation. What happened to it? And if it can have observable consequences,
why cannot the massive, negative norm particles?

Considering the pretensions of S-matrix chauvinists they should be required to 
explain in detail how to infer known gravitational phenomena from the results of
asymptotic scattering experiments. For example, how does one predict the tides?
What scattering experiment describes the primordial power spectra? And it is not
fair reconstructing a local field theory from the perturbative S-matrix, then 
using this. S-matrix chauvinists have denied the reality of local fields in 
order to avoid acknowledging the problems associated with negative norms. Let 
them live by the rules they have proclaimed.

\subsection{An Explicit Counter-Example}

Two years ago I stumbled upon a nonlinear sigma model with a trivial S-matrix 
which nonetheless shows interesting evolution of its background and of single 
particle kinematics \cite{Miao:2021gic},
\begin{equation}
\mathcal{L} = -\frac12 \partial_{\mu} A \partial_{\nu} A g^{\mu\nu} \sqrt{-g}
- \frac12 \Bigl(1 + \frac{\lambda}{2} A\Bigr)^2 \partial_{\mu} B \partial_{\nu} B
g^{\mu\nu} \sqrt{-g} \; . \label{ABmodel}
\end{equation}
This model can be reduced to a theory of two free scalars by making a local,
invertible field redefinition,\footnote{I thank Arkady Tseytlin for this
observation.}
\begin{eqnarray}
X &\!\!\! \equiv \!\!\!& \frac{2}{\lambda} \Bigl(1 + \frac{\lambda}{2} A\Bigr)
\cos\Bigl( \frac{\lambda}{2} B\Bigr) \; , \label{redef1} \\
Y &\!\!\! \equiv \!\!\!& \frac{2}{\lambda} \Bigl(1 + \frac{\lambda}{2} A\Bigr)
\sin\Bigl( \frac{\lambda}{2} B\Bigr) \; . \label{redef2}
\end{eqnarray}
Hence its S-matrix is unity by Borchers Theorem \cite{Borchers:1960}, and an
S-matrix chauvinist would be required to dismiss the model as completely
uninteresting.

In spite of having a unit S-matrix, the model (\ref{ABmodel}) is still
interesting on de Sitter background,
\begin{equation}
ds^2 = -c^2 dt^2 + e^{2 H t} d\vec{x} \!\cdot\! d\vec{x} \; . \label{deSitter}
\end{equation}
The Lagrangian (\ref{ABmodel}) is invariant under $B \rightarrow -B$, which 
precludes the field $B(x)$ from developing an expectation value. However, no 
symmetry protects $A(x)$, and explicit computations at 1-loop \cite{Miao:2021gic} 
and 2-loop \cite{Woodard:2023rqo} reveal a fascinating secular growth,
\begin{equation}
\Bigl\langle \Omega \Bigl\vert A(x) \Bigr\vert \Omega \Bigr\rangle = 
\frac{\lambda H^2 \!\cdot\! H t}{2^4 \pi^2} + \frac{\lambda^3 H^4 (Ht)^2}{
2^{10} \pi^4} + O(\lambda^5) \; . \label{VEVA}
\end{equation}
A 1-loop computation of the self-mass also shows that $A$ particles
develop a mass \cite{Miao:2021gic},
\begin{equation}
m^2_A = \frac{3 \lambda^2 H^4}{32 \pi^2} + O(\lambda^4) \; . \label{massA}
\end{equation}
Note that the absence of scattering {\it between} particle states in no way 
precludes changes in single particle kinematics, nor does it prevent changes
in the background. Neither (\ref{VEVA}), nor (\ref{massA}), is derivable from
the trivial S-matrix, and it requires some hardihood to maintain that these
results are uninteresting.

The model (\ref{ABmodel}) fascinated me and my collaborators (Shun-Pei Miao 
and Nick Tsamis) because it taught us how to sum up the potentially large 
factors of $H t$ engendered by loops of gravitons on de Sitter background
\cite{Glavan:2013jca,Wang:2014tza,Miao:2006gj,Glavan:2021adm,Tan:2021lza,
Tan:2022xpn}. The method is to construct a curvature-induced effective 
potential by integrating out differentiated $B$ fields from the $A$ field 
equation in a constant $A$ background \cite{Miao:2021gic}, 
\begin{eqnarray}
\frac{\delta S[A,B]}{\delta A} &\!\!\! = \!\!\!& \partial_{\mu} \Bigl[
\sqrt{-g} \, g^{\mu\nu} \partial_{\nu} A\Bigr] - \frac{\lambda}{2} \Bigl(1 +
\frac{\lambda}{2} A\Bigr) \partial_{\mu} B \partial_{\nu} B g^{\mu\nu} 
\sqrt{-g} \; , \qquad \\
&\!\!\! \longrightarrow \!\!\!& \partial_{\mu} \Bigl[\sqrt{-g} \, g^{\mu\nu} 
\partial_{\nu} A\Bigr] - \frac{\frac{\lambda}{2} \sqrt{-g} \, g^{\mu\nu}
\partial_{\mu} \partial'_{\nu} i\Delta(x;x') \vert_{x'=x}}{1 + \frac{\lambda}{2}
A} \; . \qquad \label{stocheqn}
\end{eqnarray}
Dimensional regularization on de Sitter implies \cite{Onemli:2002hr,
Onemli:2004mb},
\begin{equation}
g^{\mu\nu} \partial_{\mu} \partial'_{\nu} i\Delta(x;x') \vert_{x'=x} =
-\frac{H^D \Gamma(D)}{(4\pi)^{\frac{D}2} \Gamma(\frac{D}2)} \longrightarrow
-\frac{3 H^4}{8\pi^2} \; .
\end{equation} 
The result is a scalar potential model for $A$ with $V_{\rm eff}(A) = -
\frac{3 H^4}{8\pi^2} \ln\vert 1 + \frac{\lambda}{2} A\vert$.\footnote{Note
that this immediately explains the mass (\ref{massA}).} Starobinsky has 
shown that such models are equivalent, at leading order in $H t$ for each 
loop, to a stochastic random field $\mathcal{A}(t,\vec{x})$ which obeys
the Langevin equation \cite{Starobinsky:1986fx,Starobinsky:1994bd},
\begin{equation}
3 H \Bigl[ \dot{\mathcal{A}} - \dot{\mathcal{A}}_0\Bigr] = -
V'_{\rm eff}(\mathcal{A}) = \frac{\frac{3 \lambda H^3}{16 \pi^2}}{1 \!+\!
\frac{\lambda}{2} \mathcal{A}} \; . \label{Langevin}
\end{equation}
The stochastic jitter is supplied by the infrared-truncation of the free
field mode sum,
\begin{equation}
\mathcal{A}_0(t,\vec{x}) = \int \!\! \frac{d^3k}{(2\pi)^3} 
\frac{\theta(k\!-\!H) \theta(e^{Ht} H \!-\! k) H}{\sqrt{2 k^3}} \Bigl[
a(\vec{k}) e^{i \vec{k} \cdot \vec{x}} + a^{\dagger}(\vec{k}) e^{-i \vec{k}
\cdot \vec{x}} \Bigr] \; . \label{jitter}
\end{equation}
If we ignore the stochastic jitter, equation (\ref{Langevin}) can be solved
exactly. Because it is easier to fluctuate down the potential than up, the
effect of the stochastic jitter is just to accelerate the roll-down evident
in this solution,
\begin{equation}
\Bigl\langle \Omega \Bigl\vert \mathcal{A}(t,\vec{x}) \Bigr\vert \Omega
\Bigr\rangle = \frac{2}{\lambda} \Biggl[ \sqrt{1 + \frac{\lambda^2 H^2
\!\cdot\! H t}{16 \pi^2} } - 1 \Biggr] + \Bigl({\rm Stochastic\ Acceleration}
\Bigr) \; . \label{LLogsolution}
\end{equation}
Note that we can therefore sum up the secular growth factors to determine 
what becomes of them after perturbation theory breaks down. Note also that
this model continues to evolve, even at arbitrarily late time.

A final point is that the knowledge we gained had nothing to do with the
model being reducible to a free theory. To see this it suffices to make a
slight change in the Lagrangian (\ref{ABmodel}) \cite{Litos:2023},
\begin{equation}
\mathcal{L} = -\frac12 \partial_{\mu} A \partial_{\nu} A g^{\mu\nu} \sqrt{-g}
- \frac12 \Bigl(1 + \frac{\lambda}{4} A\Bigr)^4 \partial_{\mu} B \partial_{\nu} B
g^{\mu\nu} \sqrt{-g} \; . \label{newAB}
\end{equation}
The field space metric of this model has nonzero curvature so it cannot be 
reduced to a free theory. Yet perturbative computations show the same secular
growth factors. Integrating out the differentiated $B$ fields results in a
very similar effective potential, $V_{\rm eff}(A) = -\frac{3 H^4}{4\pi^2} 
\ln\vert 1 + \frac{\lambda}{4} A\vert$, which gives rise to a very similar
solution of the Langevin equation,
\begin{equation}
\Bigl\langle \Omega \Bigl\vert \mathcal{A}(t,\vec{x}) \Bigr\vert \Omega
\Bigr\rangle = \frac{4}{\lambda} \Biggl[ \sqrt{1 + \frac{\lambda^2 H^2
\!\cdot\! H t}{32 \pi^2} } - 1 \Biggr] + \Bigl({\rm Stochastic\ Acceleration}
\Bigr) \; . \label{newLLog}
\end{equation}

\section{The Road Less Traveled}

It seems to me that we who do quantum gravity may be losing our way in the
search for an ultraviolet completion to general relativity. I have already
explained what is wrong with rehabilitating the Weyl counterterm, and with
over-reliance on the S-matrix. But without regard to the viability of these
undertakings, I question the goal itself. We do not now infer gravitational
phenomena --- classical or quantum --- through asymptotic scattering 
experiments, and there is little chance that we will ever do so. The same
comment applies to understanding the last stages of black hole evaporation.
I worry that the effort to develop these aspects of the theory is diverting 
attention from things we can be, and should be doing. I also suspect we are
too quick to accept the verdict of perturbation theory on quantum general
relativity.

\subsection{What We Can Do with Quantum Gravity}

Different people seek different things from quantum gravity. Two possibilities
which fascinate me are:
\begin{itemize}
\item{Blurring of the light-cone; and}
\item{Interactions with perturbations generated by primordial 
inflation.}
\end{itemize}
Note that both of these phenomena can be studied using general relativity as
a low energy effective field theory \cite{Donoghue:1994dn,Donoghue:1995cz,
Burgess:2003jk,Donoghue:2012zc,Donoghue:2022eay}. Neither of them relies on
asymptotic scattering theory.

\subsubsection{Blurring of the Light-Cone}

In general relativity it is the metric tensor which sets the light-cone.
Because the metric tensor is a quantum operator, the light-cone must
fluctuate inside and outside its average value. This was recognized way
back in the 1950's \cite{Pauli:1956,Deser:1957zz}, and thoughtful 
researchers have considered it from time to time \cite{Ford:2005rs}.
Now that it is no longer forbidden to consider local fields (see 
Section 3.1) it is possible to study the phenomenon directly by following
the propagation of a disturbance under the impact of quantum gravitational 
fluctuations. 

Suppose we solve the quantum gravitationally corrected Maxwell equation
(\ref{finalQMax}) with a point dipole which is created at the origin at
$t = 0$,
\begin{equation}
J^0(t,\vec{x}) = -\theta(t) \vec{p} \!\cdot\! \vec{\nabla} \delta^3(\vec{x})
\qquad , \qquad J^i(t,\vec{x}) = p^i \delta(t) \delta^3(\vec{x}) \; .
\label{dipole}
\end{equation}
The resulting magnetic field consists of an outward pulse 
\cite{Leonard:2012fs},
\begin{equation}
F_{ij}(t,\vec{x}) = \Bigl(p_i \partial_j - p_j \partial_i\Bigr) 
\frac{\theta(\Delta t)}{2\pi} \Bigl[1 - \frac{10 \hbar G}{3 \pi c^3} 
\frac{\partial}{\partial r^2} + O(G^2) \Bigr] \delta(r^2 - c^2 t^2) \; .
\label{pulse}
\end{equation}
One can see that the derivatives push the response infinitesimally outside
the light-cone. A slightly superluminal pulse has also been reported for 
the gravitational response to a transient source in the presence of the
quantum fluctuations of a scalar field \cite{Marunovic:2011zw}.

The degree of superluminality present in expression (\ref{pulse}) is
very small; the pulse gets no more than a Planck length outside the 
light-cone. But it does get outside, and this makes one wonder how the
effect could be strengthened. Perhaps a sufficiently advanced technology 
might permit us to build starships which ride the outward fluctuations of 
the light-cone the same way that a surfer rides a wave? A more immediate
issue is whether or not it might be possible to collect observational
evidence of superluminal propagation from astrophysical pulses traveling
enormous distances. 

\subsubsection{Modifying Physics with Inflationary Gravitons}

The accelerated expansion of inflation modifies the energy-time uncertainty
principle so that any sufficiently long wavelength, massless virtual particle 
can persist forever. Most massless particles are classically conformally
invariant, which causes an exponential redshift in the rate at which they 
emerge from the vacuum, however, gravitons and massless, minimally coupled 
scalars are produced copiously. On de Sitter background with Hubble constant
$H$, the occupation number of gravitons with one of the two possible 
polarizations, and a single wave vector $\vec{k}$ out of the {\it infinite}
possibilities, is,
\begin{equation}
N(t,\vec{k}) = \Bigl[ \frac{H e^{Ht}}{2 c k}\Bigr]^2 \; . \label{Ngrav}
\end{equation}
These gravitons interact with themselves and with other particles, and the
fact that their numbers grow endows quantum corrections with temporal and
sometimes spatial variation. 

One can compute the vacuum polarization $-i[\mbox{}^{\mu} \Pi^{\nu}](x;x')$ 
from a loop of gravitons on de Sitter background \cite{Leonard:2013xsa}, 
and then use it in the quantum-corrected Maxwell equation (\ref{QMax}). The 
results for the Coulomb potential of a static point charge $Q$ 
\cite{Glavan:2013jca}, and for the electric field strength of a spatial
plane wave photon \cite{Wang:2014tza} are,
\begin{eqnarray}
\Phi(t,r) &\!\!\! = \!\!\!& \frac{Q e^{-Ht}}{4\pi \epsilon_0 r} \Biggl\{1 +
\frac{2 \hbar G e^{-2Ht}}{3\pi c^3 r^2} + \frac{2 \hbar G H^2}{\pi c^5} 
\ln\Bigl[ \frac{e^{Ht} H r}{c}\Bigr] + \ldots \Biggr\} \; , \label{Coulomb} \\
F^{0i}(t,\vec{x}) &\!\!\! = \!\!\!& F^{0i}_{\rm tree}(t,\vec{x}) \Biggl\{1
+ \frac{2 \hbar G H^2}{\pi c^5} \ln\Bigl[ e^{Ht}\Bigr] + \ldots \Biggr\} \; .
\label{FStrength}
\end{eqnarray}
The $G/r^2$ correction in (\ref{Coulomb}) represents the de Sitter extension 
of a flat space effect that has long been known \cite{Radkowski:1970}. In 
contrast, the terms proportional to $G H^2$ in (\ref{Coulomb}-\ref{FStrength})
derive from inflationary gravitons, and both grow with time. Similar 1-loop 
graviton effects have been reported for the field strength of fermions 
\cite{Miao:2006gj}, for the exchange potential of a massless, minimally coupled
scalar \cite{Glavan:2021adm}, for gravitational radiation \cite{Tan:2021lza}, 
and for the gravitational response to a point mass \cite{Tan:2022xpn}.

A fascinating aspect of results such as (\ref{Coulomb}-\ref{FStrength})
is that they grow stronger the longer the de Sitter expansion persists.
This must eventually overwhelm even the smallest loop-counting parameter,
at which point perturbation theory breaks down. The nonlinear sigma model
(\ref{ABmodel}) described in Section 3.3 was introduced in order to develop
a method for evolving beyond the breakdown of perturbation theory. The 
answer \cite{Miao:2021gic} combines a variant of Starobinsky's stochastic 
formalism \cite{Starobinsky:1986fx,Starobinsky:1994bd}, based on 
curvature-dependent effective potentials, with a variant of the 
renormalization group, based on the subset of counterterms which can be 
viewed as curvature-dependent renormalizations of parameters in the bare 
Lagrangian. It seems as if the technique can be applied to quantum gravity
\cite{Glavan:2021adm}. Further, the technique can be implemented for a
general cosmological background which has experienced primordial inflation
\cite{Kasdagli:2023nzj}, and significant effects persist to arbitrarily
late times \cite{Woodard:2023cqi}. Perhaps this can answer the three largest 
questions of cosmology:
\begin{itemize}
\item{What caused primordial inflation?}
\item{What caused the current phase of cosmic acceleration?}
\item{What is responsible for the phenomena ascribed to dark matter?}
\end{itemize}

\subsection{Perhaps It's Just GR}

Data from Earthbound laboratories \cite{Will:2014kxa}, all the way to the
dizzying scales of primordial inflation \cite{Planck:2018nkj}, suggest that 
gravity should be based on a local, invariant theory of a real-valued metric 
existing on a real-valued spacetime. However, the only stable extensions of 
general relativity are $f(R)$ models, which are not perturbatively 
renormalizable. This means something has to give. I have argued that 
rehabilitating the Weyl counterterm is not viable. Abandoning invariance 
allows one to employ higher spatial derivatives, without the problematic 
higher time derivatives \cite{Horava:2008ih,Horava:2009uw,Mukohyama:2010xz}. 
This suffices for renormalizability \cite{Barvinsky:2015kil}, but of course 
leaves the problem of recovering macroscopic invariance. Then again, gravity 
may not be fundamentally based on a metric \cite{Carlip:2012wa,
Verlinde:2016toy,Hossenfelder:2017eoh}, but one must then explain why the
metric-based theory seems to apply up to $10^{14}~{\rm GeV}$ during
primordial inflation. I want here to consider the other possibility: that
the problem lies with perturbation theory.

If quantum general relativity makes sense nonperturbatively one might be 
able to define it by taking the continuum limit of a numerical lattice 
calculation \cite{Ambjorn:2012jv,Ambjorn:2020rcn}. In view of how long it 
took to obtain good results in the vastly simpler problem of lattice QCD, 
it is not surprising that progress exploring this possibility has been slow. 
A simpler, not inconsistent route might be to develop a new perturbative 
expansion which incorporates logarithms and fractional powers. 

\begin{figure}[ht]
\centering
\includegraphics[width = 8cm]{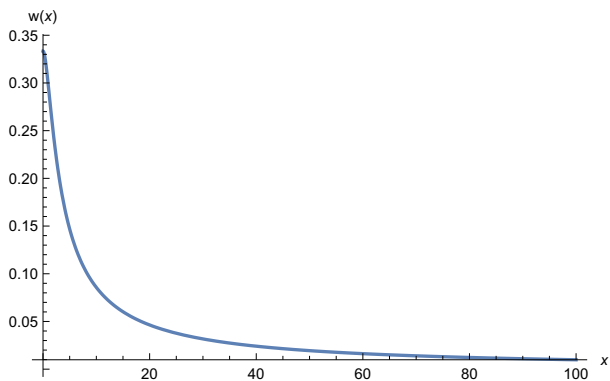}
\caption{\footnotesize Equation of state $w(x)$ of a particle of mass $m$ at 
temperature $T$, where $x \equiv m c^2/k_B T$.}
\label{stateplot}
\end{figure}

An example of some relevance to cosmology is the equation of state of a 
particle of mass $m$ at temperature $T$,
\begin{equation}
w(x) = \frac1{3 + x - x \frac{\partial}{\partial x} \ln[f(x)]} \qquad ,
\qquad x \equiv \frac{m c^2}{k_B T} \; , \label{wdef}
\end{equation}
where the function $f(x)$ is,
\begin{equation}
f(x) = \int_{0}^{\infty} \!\!\!\! dt \, (t + x) \sqrt{t^2 + 2 x t} \,
e^{-t} \; . \label{fdef}
\end{equation}
Figure~\ref{stateplot} shows that (\ref{wdef}) interpolates smoothly between 
the ultrarelativistic limit of $w(0) = \frac13$ to the nonrelativistic limit 
of $w(\infty) = 0$.
\noindent However, the series expansion in powers of $x$ is a little tricky.
Expanding the integrand would lead one to expect that the coefficient of the 
$x^3$ term vanishes, and that the $x^4$ term diverges,
\begin{equation}
f(x) = \int_{0}^{\infty} \!\!\!\! dt \Bigl[ t^2 + 2 t x + \frac12 x^2
+ 0 \cdot \frac{x^3}{t} - \frac{x^4}{8 t^2} + \ldots\Bigr] e^{-t} \; .
\label{fseries}
\end{equation}
Both expectations are wrong: the $x^3$ term has a nonzero coefficient
and the next order term is perfectly finite and proportional to $x^4 
\ln(x)$,
\begin{equation}
f(x) = 2 + 2 x + \frac12 x^2 - \frac16 x^3 - \frac18 x^4 \ln(x) + O(x^4)
\; . \label{trueseries}
\end{equation}
If such a simple system can develop logarithms in its series expansion,
why can this not happen for quantum gravity?

The breakdown of perturbation theory might be tied to the presence of 
divergences. In a renormalizable theory these divergences are canceled
by counterterms, so that the sum of a primitive diagram and the associated
counterterm can remain perturbatively small. However, that is precisely what
fails in quantum general relativity. Perhaps finiteness comes instead from the 
gravitational response to divergences? Dvali's work on classicalization 
\cite{Dvali:2010jz,Dvali:2014ila} may be relevant.

An old classical calculation by Arnowitt, Deser and Misner provides a thought
provoking example \cite{Arnowitt:1960zz}. They considered the mass of a point 
particle with bare mass $M_0$ and charge $Q$, regulated as a spherical shell 
of radius $R$. Although ADM solved the full general relativistic constraints, 
their result can be understood using a simple model they devised,
\begin{equation}
M c^2 = M_0 c^2 \!+\! \frac{Q^2}{8\pi \epsilon_0 R} \!-\! \frac{G M^2}{2 R}
\Longrightarrow M(R) = \frac{c^2 R}{G} \Biggl[ \sqrt{1 \!+\! \frac{2 G M_0}{R c^2}
\!+\! \frac{G Q^2}{4 \pi \epsilon_0 R^2 c^4} } - 1 \Biggr] . \label{ADM1}
\end{equation}
Note that the unregulated limit is finite and, interestingly, independent of
the bare mass,
\begin{equation}
\lim_{R \rightarrow 0} M(R) = \sqrt{ \frac{Q^2}{4\pi \epsilon_0 G}} \; . \label{ADM2}
\end{equation}
This is not at all what one finds using perturbation theory. Expanding the
square root in expression (\ref{ADM1}) reveals an escalating series of ever
higher divergences,
\begin{eqnarray}
\lefteqn{M(R) = \Bigl( M_0 \!+\! \frac{Q^2}{8\pi \epsilon_0 R c^2}\Bigr) \Biggl\{1 
- \frac14 \Bigl(\frac{2 G M_0}{R c^2} \!+\! \frac{G Q^2}{4\pi \epsilon_0 R^2 c^4}\Bigr) 
} \nonumber \\
& & \hspace{1cm} + \frac18 \Bigl(\frac{2 G M_0}{R c^2} \!+\! \frac{G Q^2}{4\pi 
\epsilon_0 R^2 c^4} \Bigr)^2 -\frac{5}{64} \Bigl(\frac{2 G M_0}{R c^2} \!+\! 
\frac{G Q^2}{4\pi \epsilon_0 R^2 c^4} \Bigr)^3 + \ldots\Biggr\} . \qquad \label{ADM3}
\end{eqnarray}
Of course perturbation theory is not valid when the expansion parameter becomes
infinite. Perhaps the same problem invalidates the use of perturbation theory in
quantum general relativity, which would show similar cancellations if only we could
devise a better approximation scheme? Several studies have searched for one without
success \cite{Woodard:1997dm,Casadio:2009eh,Casadio:2009jc,Mora:2011jj}, but the
amount of effort expended is minuscule compared when compared with the recurrent 
attempts to rehabilitate the Weyl counterterm. 

Cosmology offers an example of the gravitational constraints almost completely
canceling scalar perturbations during primordial inflation. Suppose the cosmological 
scale factor is $a(t)$. Two of its derivatives are the Hubble parameter $H(t)$ and 
the first slow roll parameter $\epsilon(t)$,
\begin{equation}
H(t) \equiv \frac{\dot{a}}{a} \qquad , \qquad \epsilon(t) \equiv -\frac{\dot{H}}{H^2}
\; . \label{cparams}
\end{equation}
Single scalar inflation consists of general relativity plus a minimally coupled
inflaton $\varphi$ whose slow roll down its potential $V(\varphi)$ provides the
stress-energy of inflation,
\begin{equation}
\mathcal{L} = \frac{c^4 R \sqrt{-g}}{16 \pi G} - \frac12 \partial_{\mu} \varphi
\partial_{\nu} \varphi g^{\mu\nu} \sqrt{-g} - V(\varphi) \sqrt{-g} \; . 
\label{infL}
\end{equation}
The scalar background and its potential are related to the parameters 
(\ref{cparams}),
\begin{equation}
\dot{\varphi}_0^2 = -\frac{c^4 \dot{H}}{4\pi G} \qquad , \qquad V(\varphi_0) =
\frac{c^2 (\dot{H} \!+\! 3 H^2)}{8\pi G} \; . \label{infback}
\end{equation}
It is usual to employ the ADM parameterization for the metric \cite{Arnowitt:1959ah},
\begin{equation}
ds^2 = -N^2 c^2 dt^2 + h_{ij} \Bigl(dx^i \!-\! N^i c dt\Bigr) \Bigl(dx^j \!-\! N^j
c dt\Bigr) \; . \label{ADMmetric}
\end{equation}
The 3-metric is written in terms of a component $\zeta$ and a traceless part
$\chi_{ij}$,
\begin{equation}
h_{ij}(t,\vec{x}) \equiv a^2(t) \!\times\! e^{2 \zeta(t,\vec{x})} \!\times\! 
\Bigl[e^{\chi(t,\vec{x})}\Bigr]_{ij} \qquad , \qquad \chi_{ii}(t,\vec{x}) = 0 \; . 
\label{infmetric}
\end{equation}
Instead of the lapse $N(t,\vec{x})$ and shift $N^i(t,\vec{x})$ being gauge choices, 
the gauge conditions are,
\begin{equation}
\varphi(t,\vec{x}) = \varphi_0(t) \qquad , \qquad \partial_j \chi_{ij}(t,\vec{x}) 
= 0 \; . \label{gauge}
\end{equation}
The lapse and shift are instead determined by solving the constraints. The funny
thing is, that doing so leads to the almost total cancellation of the scalar 
perturbation $\zeta(t,\vec{x})$,
\begin{equation}
\mathcal{L} \longrightarrow \frac{c^4 a^3 \epsilon}{8\pi G} \Bigl[
\frac{\dot{\zeta}^2}{c^2} - \frac{\partial_k \zeta \partial_k \zeta}{a^2} \Bigr] 
+ \frac{c^4 a^3}{64 \pi G} \Bigl[ \frac{\dot{\chi}_{ij} \dot{\chi}_{ij}}{c^2} -
\frac{\partial_k \chi_{ij} \partial_k \chi_{ij}}{a^2} \Bigr] + \Bigl({\rm 
Interactions}\Bigr) \; . \label{constL}
\end{equation}

Note from (\ref{infL}) that the scalar perturbation had unit strength before
imposing the constraints, even after gauge fixing (\ref{gauge}). The gravitational
constraints have almost completely erased it at the quadratic level (\ref{constL}).
Detailed calculations of the constrained interactions \cite{Maldacena:2002vr,
Seery:2006vu,Jarnhus:2007ia,Xue:2012wi} show that each additional one or two powers 
of the scalar perturbation leads to suppression by an additional factor of 
$\epsilon$. To understand how significant the cancellation is, recall that 
approximate values for the scalar and tensor power spectra at wave number $k$ can 
be written in terms of the geometrical parameters (\ref{cparams}) evaluated at 
the horizon crossing time $t_k$ such that $c k = H(t_k) a(t_k)$,
\begin{equation}
\Delta^2_{\mathcal{R}}(k) \simeq \frac{\hbar G H^2(t_k)}{\pi c^5 \epsilon(t_k)}
\qquad , \qquad \Delta^2_{h}(k) \simeq \frac{16 \hbar G H^2(t_k)}{\pi c^5} \; .
\label{pspectra}
\end{equation}
The fact that the scalar power spectrum has been observed, to three significant 
figures and over a range of about 8 e-foldings \cite{Planck:2018nkj}, while the 
tensor power spectrum has yet to be resolved \cite{BICEP:2021xfz}, means that 
the first slow roll parameter is very small, $\epsilon \ltwid 0.0023$.

\section{Conclusions}

The gravitational data we currently possess are all either classical 
\cite{Will:2014kxa} or else semi-classical \cite{Planck:2018nkj}. There 
is no point to defining a theory of ``quantum gravity'' which fails to 
explain these data. This leads to the {\it Quantum Gravitational 
Correspondence Principle} that the classical limit of any proposal for 
``quantum gravity'' must consist of a local, invariant theory of a 
real-valued metric on a real-valued spacetime. The Correspondence Principle
poses an obstacle for modifications of general relativity because we have a 
complete catalog of such theories and, with the exception of $f(R)$ models, 
all of them are subject classically to a virulent kinetic energy instability 
which prevents them even having a finite decay rate. Alternate quantization 
schemes which claim to avoid this instability fail to obey the Correspondence 
Principle, and obligate their advocates to explain, in some detail, how they
recover the vast body of gravitational phenomena.

These considerations apply to the dimension four Lagrangian (\ref{dim4L}) 
which is the focus of many efforts to quantize gravity. In Section 2 I 
explain how the procedure of regarding negative energy creation operators as 
positive energy annihilation operators produces the renormalizable, negative 
norm quantization. This quantization massively violates the Correspondence 
Principle. One manifestation of this violation is that the square of the time 
derivative of the geodesic length-squared along a spacelike interval is 
negative ($[ \int \dot{g}_{ij} dx^i dx^j]^2 < 0$), even over macroscopic 
separations. This is not some tiny, Planck-suppressed effect; it occurs at 
order one in the classical limit, and it is totally unacceptable. Note that 
the negative sign cannot be avoided by careful mathematics, or by further 
tinkering with the field theory, because it is required for renormalizability.

People who seek to rehabilitate the Weyl counterterm in (\ref{dim4L}) argue 
that local fields such as $\dot{g}_{ij}(t,\vec{x})$ are not observable, that 
the theory can only be defined by the perturbative S-matrix on flat space 
background, and that the negative norm states are no problem because they
not even present in the asymptotic scattering space. Section 3 criticizes 
this view, pointing out that it is perfectly reasonable to study 
local fields, and that {\it all} current gravitational data --- including even 
the primordial power spectra --- are analyzed in precisely this manner. I 
challenge S-matrix extremists to explain these data using asymptotic scattering 
theory, without recourse to local fields. It is also worth noting that the 
quantum gravitational S-matrix fails to exist on flat space background owing 
to the infrared problem, and that even inclusive rates and cross sections are 
unlikely to exist or be observable in cosmology. Note that the problem with 
observability is not some subtle issue which might be circumvented with 
careful mathematics; it is rather that causality and spacetime expansion 
preclude performing the required measurements.

The really crushing argument against S-matrix chauvinism comes in Section 3.3 
where I present a nonlinear sigma model (\ref{ABmodel}) on de Sitter background 
which is reducible to a free theory by a local, invertible field redefinition
(\ref{redef1}-\ref{redef2}). This means that its S-matrix is unity and, if we 
adhere to S-matrix chauvinism, nothing interesting happens in the theory.
But the scalar background still shows a fascinating evolution (\ref{VEVA}), 
leading to corresponding changes in the masses (\ref{massA}) of single 
particles, all despite the absence of scattering.

It seems to me that the attention devoted to high energy scattering theory and 
to black hole evaporation have led us away from the testable and exciting things 
quantum gravity can do for us in the context of low energy effective field 
theory \cite{Donoghue:1994dn,Donoghue:1995cz,Burgess:2003jk,Donoghue:2012zc,
Donoghue:2017ovt,Donoghue:2022eay}. Section 4 reviews my own favorites. I also
discuss the possibility that we should take seriously the difficulty of modifying
general relativity and focus instead on the inappropriate application of 
perturbation theory such as expression (\ref{ADM3}) when the actual series
expansion includes fractional powers or logarithms of $G$.

Finally, I should comment that either possibility for the viability of the 
dimension four Lagrangian (\ref{dim4L}) is problematic for the program of 
Asymptotic Safety \cite{Niedermaier:2006wt,Benedetti:2009rx,Falls:2018ylp}. 
Either the Weyl counterterm is permitted or it is not. If it is allowed then 
(\ref{dim4L}) is perturbatively renormalizable, and we have the ultraviolet 
completion of general relativity, without the need for any higher 
counterterms. On the other hand, if the Weyl counterterm is not permitted 
then most of the higher counterterms are also forbidden, and those from 
$f(R)$ extensions will not suffice to absorb all divergences. Either way, 
motivation is lacking to search for fixed points of the infinite collection 
of higher counterterms. If we instead treat the higher counterterms 
perturbatively, in the sense of low energy effective field theory 
\cite{Donoghue:1994dn,Donoghue:1995cz,Burgess:2003jk,Donoghue:2012zc,
Donoghue:2017ovt,Donoghue:2022eay}, then their unknown coefficients pose
no problem to predictability, in any existing or projected data sets, as 
long as their coefficients are of order one in Planck units. 

That last caveat, about the unknown coefficients being of order one in Planck
units, is significant because violations have been proposed for $f(R)$ models,
which are the sole allowed extensions of general relativity. In particular, 
permitting the $R^2$ coefficient $\alpha$ to be as large as $10^{10}$ provides 
a model of inflation \cite{Starobinsky:1980te} which agrees well with existing 
data \cite{Planck:2018jri}. $f(R)$ models can also explain late time 
acceleration within observational limits \cite{Starobinsky:2007hu}. It would 
be a huge triumph for Asymptotic Safety to justify these models. 

\vskip 1cm

\centerline{\bf Acknowledgements}

I am grateful for civil discussion and correspondence on this subject with 
J. F. Donoghue, B. Holdom and P. D. Mannheim. This work was supported by NSF 
grant PHY-2207514 and by the Institute for Fundamental Theory at the 
University of Florida.


\begin{thebibliography}{99}

%\cite{Will:2014kxa}
\bibitem{Will:2014kxa}
C.~M.~Will,
%``The Confrontation between General Relativity and Experiment,''
Living Rev. Rel. \textbf{17}, 4 (2014)
doi:10.12942/lrr-2014-4
[arXiv:1403.7377 [gr-qc]].
%1977 citations counted in INSPIRE as of 24 May 2023

%\cite{Mukhanov:1981xt}
\bibitem{Mukhanov:1981xt}
V.~F.~Mukhanov and G.~V.~Chibisov,
%``Quantum Fluctuations and a Nonsingular Universe,''
JETP Lett. \textbf{33}, 532-535 (1981)
%1949 citations counted in INSPIRE as of 29 May 2023

%\cite{Planck:2018nkj}
\bibitem{Planck:2018nkj}
N.~Aghanim \textit{et al.} [Planck],
%``Planck 2018 results. I. Overview and the cosmological legacy of Planck,''
Astron. Astrophys. \textbf{641}, A1 (2020)
doi:10.1051/0004-6361/201833880
[arXiv:1807.06205 [astro-ph.CO]].
%1236 citations counted in INSPIRE as of 29 May 2023

%\cite{Loeb:2003ya}
\bibitem{Loeb:2003ya}
A.~Loeb and M.~Zaldarriaga,
%``Measuring the small - scale power spectrum of cosmic density fluctuations through 21 cm tomography prior to the epoch of structure formation,''
Phys. Rev. Lett. \textbf{92}, 211301 (2004)
doi:10.1103/PhysRevLett.92.211301
[arXiv:astro-ph/0312134 [astro-ph]].
%301 citations counted in INSPIRE as of 29 May 2023

%\cite{Furlanetto:2006jb}
\bibitem{Furlanetto:2006jb}
S.~Furlanetto, S.~P.~Oh and F.~Briggs,
%``Cosmology at Low Frequencies: The 21 cm Transition and the High-Redshift Universe,''
Phys. Rept. \textbf{433}, 181-301 (2006)
doi:10.1016/j.physrep.2006.08.002
[arXiv:astro-ph/0608032 [astro-ph]].
%1000 citations counted in INSPIRE as of 29 May 2023

%\cite{Masui:2010cz}
\bibitem{Masui:2010cz}
K.~W.~Masui and U.~L.~Pen,
%``Primordial gravity wave fossils and their use in testing inflation,''
Phys. Rev. Lett. \textbf{105}, 161302 (2010)
doi:10.1103/PhysRevLett.105.161302
[arXiv:1006.4181 [astro-ph.CO]].
%67 citations counted in INSPIRE as of 29 May 2023

%\cite{Starobinsky:1979ty}
\bibitem{Starobinsky:1979ty}
A.~A.~Starobinsky,
%``Spectrum of relict gravitational radiation and the early state of the universe,''
JETP Lett. \textbf{30}, 682-685 (1979)
%1865 citations counted in INSPIRE as of 29 May 2023

%\cite{BICEP:2021xfz}
\bibitem{BICEP:2021xfz}
P.~A.~R.~Ade \textit{et al.} [BICEP and Keck],
%``Improved Constraints on Primordial Gravitational Waves using Planck, WMAP, and BICEP/Keck Observations through the 2018 Observing Season,''
Phys. Rev. Lett. \textbf{127}, no.15, 151301 (2021)
doi:10.1103/PhysRevLett.127.151301
[arXiv:2110.00483 [astro-ph.CO]].
%393 citations counted in INSPIRE as of 24 May 2023

%\cite{Tan:2021lza}
\bibitem{Tan:2021lza}
L.~Tan, N.~C.~Tsamis and R.~P.~Woodard,
%``How inflationary gravitons affect gravitational radiation,''
Phil. Trans. Roy. Soc. Lond. A \textbf{380}, 0187 (2021)
doi:10.1098/rsta.2021.0187
[arXiv:2107.13905 [gr-qc]].
%7 citations counted in INSPIRE as of 29 May 2023

%\cite{Donoghue:1994dn}
\bibitem{Donoghue:1994dn}
J.~F.~Donoghue,
%``General relativity as an effective field theory: The leading quantum corrections,''
Phys. Rev. D \textbf{50}, 3874-3888 (1994)
doi:10.1103/PhysRevD.50.3874
[arXiv:gr-qc/9405057 [gr-qc]].
%1104 citations counted in INSPIRE as of 24 May 2023

%\cite{Donoghue:1995cz}
\bibitem{Donoghue:1995cz}
J.~F.~Donoghue,
%``Introduction to the effective field theory description of gravity,''
[arXiv:gr-qc/9512024 [gr-qc]].
%256 citations counted in INSPIRE as of 24 May 2023

%\cite{Burgess:2003jk}
\bibitem{Burgess:2003jk}
C.~P.~Burgess,
%``Quantum gravity in everyday life: General relativity as an effective field theory,''
Living Rev. Rel. \textbf{7}, 5-56 (2004)
doi:10.12942/lrr-2004-5
[arXiv:gr-qc/0311082 [gr-qc]].
%515 citations counted in INSPIRE as of 24 May 2023

%\cite{Donoghue:2012zc}
\bibitem{Donoghue:2012zc}
J.~F.~Donoghue,
%``The effective field theory treatment of quantum gravity,''
AIP Conf. Proc. \textbf{1483}, no.1, 73-94 (2012)
doi:10.1063/1.4756964
[arXiv:1209.3511 [gr-qc]].
%167 citations counted in INSPIRE as of 24 May 2023

%\cite{Donoghue:2017ovt}
\bibitem{Donoghue:2017ovt}
J.~Donoghue,
%``Quantum gravity as a low energy effective field theory,''
Scholarpedia \textbf{12}, no.4, 32997 (2017)
doi:10.4249/scholarpedia.32997
%21 citations counted in INSPIRE as of 29 May 2023

%\cite{Donoghue:2022eay}
\bibitem{Donoghue:2022eay}
J.~F.~Donoghue,
%``Quantum General Relativity and Effective Field Theory,''
[arXiv:2211.09902 [hep-th]].
%8 citations counted in INSPIRE as of 24 May 2023

%\cite{Stelle:1976gc}
\bibitem{Stelle:1976gc}
K.~S.~Stelle,
%``Renormalization of Higher Derivative Quantum Gravity,''
Phys. Rev. D \textbf{16}, 953-969 (1977)
doi:10.1103/PhysRevD.16.953
%2188 citations counted in INSPIRE as of 24 Apr 2022

%\cite{Starobinsky:1980te}
\bibitem{Starobinsky:1980te}
A.~A.~Starobinsky,
%``A New Type of Isotropic Cosmological Models Without Singularity,''
Phys. Lett. B \textbf{91}, 99-102 (1980)
doi:10.1016/0370-2693(80)90670-X
%6213 citations counted in INSPIRE as of 24 May 2023

%\cite{Planck:2018vyg}
\bibitem{Planck:2018vyg}
N.~Aghanim \textit{et al.} [Planck],
%``Planck 2018 results. VI. Cosmological parameters,''
Astron. Astrophys. \textbf{641}, A6 (2020)
[erratum: Astron. Astrophys. \textbf{652}, C4 (2021)]
doi:10.1051/0004-6361/201833910
[arXiv:1807.06209 [astro-ph.CO]].
%10568 citations counted in INSPIRE as of 24 May 2023

%\cite{Planck:2018jri}
\bibitem{Planck:2018jri}
Y.~Akrami \textit{et al.} [Planck],
%``Planck 2018 results. X. Constraints on inflation,''
Astron. Astrophys. \textbf{641}, A10 (2020)
doi:10.1051/0004-6361/201833887
[arXiv:1807.06211 [astro-ph.CO]].
%2442 citations counted in INSPIRE as of 24 May 2023

%\cite{Woodard:2009ns}
\bibitem{Woodard:2009ns}
R.~P.~Woodard,
%``How Far Are We from the Quantum Theory of Gravity?,''
Rept. Prog. Phys. \textbf{72}, 126002 (2009)
doi:10.1088/0034-4885/72/12/126002
[arXiv:0907.4238 [gr-qc]].
%130 citations counted in INSPIRE as of 24 May 2023

%\cite{DeWitt:1967yk}
\bibitem{DeWitt:1967yk}
B.~S.~DeWitt,
%``Quantum Theory of Gravity. 1. The Canonical Theory,''
Phys. Rev. \textbf{160}, 1113-1148 (1967)
doi:10.1103/PhysRev.160.1113
%2991 citations counted in INSPIRE as of 29 May 2023

%\cite{DeWitt:1967ub}
\bibitem{DeWitt:1967ub}
B.~S.~DeWitt,
%``Quantum Theory of Gravity. 2. The Manifestly Covariant Theory,''
Phys. Rev. \textbf{162}, 1195-1239 (1967)
doi:10.1103/PhysRev.162.1195
%1701 citations counted in INSPIRE as of 29 May 2023

%\cite{DeWitt:1967uc}
\bibitem{DeWitt:1967uc}
B.~S.~DeWitt,
%``Quantum Theory of Gravity. 3. Applications of the Covariant Theory,''
Phys. Rev. \textbf{162}, 1239-1256 (1967)
doi:10.1103/PhysRev.162.1239
%976 citations counted in INSPIRE as of 29 May 2023

%\cite{tHooft:1974toh}
\bibitem{tHooft:1974toh}
G.~'t Hooft and M.~J.~G.~Veltman,
%``One loop divergencies in the theory of gravitation,''
Ann. Inst. H. Poincare Phys. Theor. A \textbf{20}, 69-94 (1974)
%1134 citations counted in INSPIRE as of 29 May 2023

%\cite{Deser:1974zzd}
\bibitem{Deser:1974zzd}
S.~Deser and P.~van Nieuwenhuizen,
%``Nonrenormalizability of the Quantized Einstein-Maxwell System,''
Phys. Rev. Lett. \textbf{32}, 245-247 (1974)
doi:10.1103/PhysRevLett.32.245
%164 citations counted in INSPIRE as of 29 May 2023

%\cite{Deser:1974cz}
\bibitem{Deser:1974cz}
S.~Deser and P.~van Nieuwenhuizen,
%``One Loop Divergences of Quantized Einstein-Maxwell Fields,''
Phys. Rev. D \textbf{10}, 401 (1974)
doi:10.1103/PhysRevD.10.401
%532 citations counted in INSPIRE as of 29 May 2023

%\cite{Deser:1974cy}
\bibitem{Deser:1974cy}
S.~Deser and P.~van Nieuwenhuizen,
%``Nonrenormalizability of the Quantized Dirac-Einstein System,''
Phys. Rev. D \textbf{10}, 411 (1974)
doi:10.1103/PhysRevD.10.411
%384 citations counted in INSPIRE as of 29 May 2023

%\cite{Deser:1974nb}
\bibitem{Deser:1974nb}
S.~Deser, H.~S.~Tsao and P.~van Nieuwenhuizen,
%``Nonrenormalizability of Einstein Yang-Mills Interactions at the One Loop Level,''
Phys. Lett. B \textbf{50}, 491-493 (1974)
doi:10.1016/0370-2693(74)90268-8
%76 citations counted in INSPIRE as of 29 May 2023

%\cite{Deser:1974xq}
\bibitem{Deser:1974xq}
S.~Deser, H.~S.~Tsao and P.~van Nieuwenhuizen,
%``One Loop Divergences of the Einstein Yang-Mills System,''
Phys. Rev. D \textbf{10}, 3337 (1974)
doi:10.1103/PhysRevD.10.3337
%259 citations counted in INSPIRE as of 29 May 2023

%\cite{Goroff:1985sz}
\bibitem{Goroff:1985sz}
M.~H.~Goroff and A.~Sagnotti,
%``QUANTUM GRAVITY AT TWO LOOPS,''
Phys. Lett. B \textbf{160}, 81-86 (1985)
doi:10.1016/0370-2693(85)91470-4
%387 citations counted in INSPIRE as of 29 May 2023

%\cite{Goroff:1985th}
\bibitem{Goroff:1985th}
M.~H.~Goroff and A.~Sagnotti,
%``The Ultraviolet Behavior of Einstein Gravity,''
Nucl. Phys. B \textbf{266}, 709-736 (1986)
doi:10.1016/0550-3213(86)90193-8
%766 citations counted in INSPIRE as of 29 May 2023

%\cite{vandeVen:1991gw}
\bibitem{vandeVen:1991gw}
A.~E.~M.~van de Ven,
%``Two loop quantum gravity,''
Nucl. Phys. B \textbf{378}, 309-366 (1992)
doi:10.1016/0550-3213(92)90011-Y
%329 citations counted in INSPIRE as of 29 May 2023

%\cite{Tomboulis:1977jk}
\bibitem{Tomboulis:1977jk}
E.~Tomboulis,
%``1/N Expansion and Renormalization in Quantum Gravity,''
Phys. Lett. B \textbf{70}, 361-364 (1977)
doi:10.1016/0370-2693(77)90678-5
%299 citations counted in INSPIRE as of 29 May 2023

%\cite{Salam:1978fd}
\bibitem{Salam:1978fd}
A.~Salam and J.~A.~Strathdee,
%``Remarks on High-energy Stability and Renormalizability of Gravity Theory,''
Phys. Rev. D \textbf{18}, 4480 (1978)
doi:10.1103/PhysRevD.18.4480
%152 citations counted in INSPIRE as of 29 May 2023

%\cite{Antoniadis:1984kd}
\bibitem{Antoniadis:1984kd}
I.~Antoniadis and N.~C.~Tsamis,
%``On the Cosmological Constant Problem,''
Phys. Lett. B \textbf{144}, 55-60 (1984)
doi:10.1016/0370-2693(84)90175-8
%30 citations counted in INSPIRE as of 29 May 2023

%\cite{Antoniadis:1986tu}
\bibitem{Antoniadis:1986tu}
I.~Antoniadis and E.~T.~Tomboulis,
%``Gauge Invariance and Unitarity in Higher Derivative Quantum Gravity,''
Phys. Rev. D \textbf{33}, 2756 (1986)
doi:10.1103/PhysRevD.33.2756
%187 citations counted in INSPIRE as of 29 May 2023

%\cite{Johnston:1987ue}
\bibitem{Johnston:1987ue}
D.~A.~Johnston,
%``Sedentary Ghost Poles in Higher Derivative Gravity,''
Nucl. Phys. B \textbf{297}, 721-732 (1988)
doi:10.1016/0550-3213(88)90555-X
%68 citations counted in INSPIRE as of 29 May 2023

%\cite{Hawking:2001yt}
\bibitem{Hawking:2001yt}
S.~W.~Hawking and T.~Hertog,
%``Living with ghosts,''
Phys. Rev. D \textbf{65}, 103515 (2002)
doi:10.1103/PhysRevD.65.103515
[arXiv:hep-th/0107088 [hep-th]].
%252 citations counted in INSPIRE as of 29 May 2023

%\cite{Salles:2014rua}
\bibitem{Salles:2014rua}
F.~d.~Salles and I.~L.~Shapiro,
%``Do we have unitary and (super)renormalizable quantum gravity below the Planck scale?,''
Phys. Rev. D \textbf{89}, no.8, 084054 (2014)
[erratum: Phys. Rev. D \textbf{90}, no.12, 129903 (2014)]
doi:10.1103/PhysRevD.89.084054
[arXiv:1401.4583 [hep-th]].
%67 citations counted in INSPIRE as of 29 May 2023

%\cite{deOSalles:2018eon}
\bibitem{deOSalles:2018eon}
F.~de O.Salles and I.~L.~Shapiro,
%``Recent Progress in Fighting Ghosts in Quantum Gravity,''
Universe \textbf{4}, no.9, 91 (2018)
doi:10.3390/universe4090091
[arXiv:1808.09015 [gr-qc]].
%10 citations counted in INSPIRE as of 29 May 2023

%\cite{Donoghue:2019fcb}
\bibitem{Donoghue:2019fcb}
J.~F.~Donoghue and G.~Menezes,
%``Unitarity, stability and loops of unstable ghosts,''
Phys. Rev. D \textbf{100}, no.10, 105006 (2019)
doi:10.1103/PhysRevD.100.105006
[arXiv:1908.02416 [hep-th]].
%88 citations counted in INSPIRE as of 29 May 2023

%\cite{Donoghue:2019ecz}
\bibitem{Donoghue:2019ecz}
J.~F.~Donoghue and G.~Menezes,
%``Arrow of Causality and Quantum Gravity,''
Phys. Rev. Lett. \textbf{123}, no.17, 171601 (2019)
doi:10.1103/PhysRevLett.123.171601
[arXiv:1908.04170 [hep-th]].
%47 citations counted in INSPIRE as of 29 May 2023

%\cite{Mannheim:2020ryw}
\bibitem{Mannheim:2020ryw}
P.~D.~Mannheim,
%``Ghost problems from Pauli\textendash{}Villars to fourth-order quantum gravity and their resolution,''
Int. J. Mod. Phys. D \textbf{29}, no.14, 2043009 (2020)
doi:10.1142/S0218271820430099
[arXiv:2004.00376 [hep-th]].
%8 citations counted in INSPIRE as of 29 May 2023

%\cite{Donoghue:2021eto}
\bibitem{Donoghue:2021eto}
J.~F.~Donoghue and G.~Menezes,
%``Ostrogradsky instability can be overcome by quantum physics,''
Phys. Rev. D \textbf{104}, no.4, 045010 (2021)
doi:10.1103/PhysRevD.104.045010
[arXiv:2105.00898 [hep-th]].
%26 citations counted in INSPIRE as of 29 May 2023

%\cite{Holdom:2021hlo}
\bibitem{Holdom:2021hlo}
B.~Holdom,
%``Ultra-Planckian scattering from a QFT for gravity,''
Phys. Rev. D \textbf{105}, no.4, 046008 (2022)
doi:10.1103/PhysRevD.105.046008
[arXiv:2107.01727 [hep-th]].
%13 citations counted in INSPIRE as of 08 Jun 2023

%\cite{Mannheim:2021oat}
\bibitem{Mannheim:2021oat}
P.~D.~Mannheim,
%``Solution to the ghost problem in higher-derivative gravity,''
Nuovo Cim. C \textbf{45}, no.2, 27 (2022)
doi:10.1393/ncc/i2022-22027-6
[arXiv:2109.12743 [hep-th]].
%4 citations counted in INSPIRE as of 29 May 2023

%\cite{Mannheim:2023mfp}
\bibitem{Mannheim:2023mfp}
P.~D.~Mannheim,
%``Normalization of the vacuum and the ultraviolet completion of Einstein gravity,''
[arXiv:2303.10827 [hep-th]].
%1 citations counted in INSPIRE as of 29 May 2023

%\cite{Lee:1969fy}
\bibitem{Lee:1969fy}
T.~D.~Lee and G.~C.~Wick,
%``Negative Metric and the Unitarity of the S Matrix,''
Nucl. Phys. B \textbf{9}, 209-243 (1969)
doi:10.1016/0550-3213(69)90098-4
%635 citations counted in INSPIRE as of 29 May 2023

%\cite{Lee:1970iw}
\bibitem{Lee:1970iw}
T.~D.~Lee and G.~C.~Wick,
%``Finite Theory of Quantum Electrodynamics,''
Phys. Rev. D \textbf{2}, 1033-1048 (1970)
doi:10.1103/PhysRevD.2.1033
%477 citations counted in INSPIRE as of 29 May 2023

%\cite{Grinstein:2007mp}
\bibitem{Grinstein:2007mp}
B.~Grinstein, D.~O'Connell and M.~B.~Wise,
%``The Lee-Wick standard model,''
Phys. Rev. D \textbf{77}, 025012 (2008)
doi:10.1103/PhysRevD.77.025012
[arXiv:0704.1845 [hep-ph]].
%231 citations counted in INSPIRE as of 29 May 2023

%\cite{Donoghue:2018lmc}
\bibitem{Donoghue:2018lmc}
J.~F.~Donoghue and G.~Menezes,
%``Massive poles in Lee-Wick quantum field theory,''
Phys. Rev. D \textbf{99}, no.6, 065017 (2019)
doi:10.1103/PhysRevD.99.065017
[arXiv:1812.03603 [hep-th]].
%15 citations counted in INSPIRE as of 29 May 2023

%\cite{Bender:2007wu}
\bibitem{Bender:2007wu}
C.~M.~Bender and P.~D.~Mannheim,
%``No-ghost theorem for the fourth-order derivative Pais-Uhlenbeck oscillator model,''
Phys. Rev. Lett. \textbf{100}, 110402 (2008)
doi:10.1103/PhysRevLett.100.110402
[arXiv:0706.0207 [hep-th]].
%290 citations counted in INSPIRE as of 29 May 2023

%\cite{Bender:2008vh}
\bibitem{Bender:2008vh}
C.~M.~Bender and P.~D.~Mannheim,
%``Giving up the ghost,''
J. Phys. A \textbf{41}, 304018 (2008)
doi:10.1088/1751-8113/41/30/304018
[arXiv:0807.2607 [hep-th]].
%48 citations counted in INSPIRE as of 29 May 2023

%\cite{Niedermaier:2006wt}
\bibitem{Niedermaier:2006wt}
M.~Niedermaier and M.~Reuter,
%``The Asymptotic Safety Scenario in Quantum Gravity,''
Living Rev. Rel. \textbf{9}, 5-173 (2006)
doi:10.12942/lrr-2006-5
%581 citations counted in INSPIRE as of 29 May 2023

%\cite{Benedetti:2009rx}
\bibitem{Benedetti:2009rx}
D.~Benedetti, P.~F.~Machado and F.~Saueressig,
%``Asymptotic safety in higher-derivative gravity,''
Mod. Phys. Lett. A \textbf{24}, 2233-2241 (2009)
doi:10.1142/S0217732309031521
[arXiv:0901.2984 [hep-th]].
%335 citations counted in INSPIRE as of 29 May 2023

%\cite{Falls:2018ylp}
\bibitem{Falls:2018ylp}
K.~G.~Falls, D.~F.~Litim and J.~Schr\"oder,
%``Aspects of asymptotic safety for quantum gravity,''
Phys. Rev. D \textbf{99}, no.12, 126015 (2019)
doi:10.1103/PhysRevD.99.126015
[arXiv:1810.08550 [gr-qc]].
%104 citations counted in INSPIRE as of 29 May 2023

%\cite{Ostrogradsky:1850fid}
\bibitem{Ostrogradsky:1850fid}
M.~Ostrogradsky,
%``M\'emoires sur les \'equations diff\'erentielles, relatives au probl\`eme des isop\'erim\`etres,''
Mem. Acad. St. Petersbourg \textbf{6}, no.4, 385-517 (1850)
%93 citations counted in INSPIRE as of 30 May 2023

%\cite{Woodard:2015zca}
\bibitem{Woodard:2015zca}
R.~P.~Woodard,
%``Ostrogradsky's theorem on Hamiltonian instability,''
Scholarpedia \textbf{10}, no.8, 32243 (2015)
doi:10.4249/scholarpedia.32243
[arXiv:1506.02210 [hep-th]].
%500 citations counted in INSPIRE as of 30 May 2023

%\cite{Woodard:2006nt}
\bibitem{Woodard:2006nt}
R.~P.~Woodard,
%``Avoiding dark energy with 1/r modifications of gravity,''
Lect. Notes Phys. \textbf{720}, 403-433 (2007)
doi:10.1007/978-3-540-71013-4\_14
[arXiv:astro-ph/0601672 [astro-ph]].
%648 citations counted in INSPIRE as of 30 May 2023

%\cite{Cline:2003gs}
\bibitem{Cline:2003gs}
J.~M.~Cline, S.~Jeon and G.~D.~Moore,
%``The Phantom menaced: Constraints on low-energy effective ghosts,''
Phys. Rev. D \textbf{70}, 043543 (2004)
doi:10.1103/PhysRevD.70.043543
[arXiv:hep-ph/0311312 [hep-ph]].
%685 citations counted in INSPIRE as of 01 Jun 2023

%\cite{Deffayet:2023wdg}
\bibitem{Deffayet:2023wdg}
C.~Deffayet, A.~Held, S.~Mukohyama and A.~Vikman,
%``Global and Local Stability for Ghosts Coupled to Positive Energy Degrees of Freedom,''
[arXiv:2305.09631 [gr-qc]].
%0 citations counted in INSPIRE as of 01 Jun 2023

%\cite{Boulware:1983td}
\bibitem{Boulware:1983td}
D.~G.~Boulware, G.~T.~Horowitz and A.~Strominger,
%``Zero Energy Theorem for Scale Invariant Gravity,''
Phys. Rev. Lett. \textbf{50}, 1726 (1983)
doi:10.1103/PhysRevLett.50.1726
%107 citations counted in INSPIRE as of 03 Jun 2023

%\cite{Mostafazadeh:2003tu}
\bibitem{Mostafazadeh:2003tu}
A.~Mostafazadeh,
%``A Critique of PT symmetric quantum mechanics,''
[arXiv:quant-ph/0310164 [quant-ph]].
%20 citations counted in INSPIRE as of 01 Jun 2023

%\cite{Mostafazadeh:2004mx}
\bibitem{Mostafazadeh:2004mx}
A.~Mostafazadeh and A.~Batal,
%``Physical aspects of pseudo-hermitian and pt-symmetric quantum mechanics,''
J. Phys. A \textbf{37}, 11645-11680 (2004)
doi:10.1088/0305-4470/37/48/009
[arXiv:quant-ph/0408132 [quant-ph]].
%145 citations counted in INSPIRE as of 01 Jun 2023

%\cite{Mostafazadeh:2010yx}
\bibitem{Mostafazadeh:2010yx}
A.~Mostafazadeh,
%``Conceptual Aspects of PT-Symmetry and Pseudo-Hermiticity: A status report,''
Phys. Scripta \textbf{82}, 038110 (2010)
doi:10.1088/0031-8949/82/03/038110
[arXiv:1008.4680 [quant-ph]].
%37 citations counted in INSPIRE as of 01 Jun 2023

%\cite{Jackson:1998nia}
\bibitem{Jackson:1998nia}
J.~D.~Jackson,
%``Classical Electrodynamics,''
Wiley, 1998,
ISBN 978-0-471-30932-1
%349 citations counted in INSPIRE as of 04 Jun 2023

%\cite{Leonard:2012fs}
\bibitem{Leonard:2012fs}
K.~E.~Leonard and R.~P.~Woodard,
%``Graviton Corrections to Maxwell's Equations,''
Phys. Rev. D \textbf{85}, 104048 (2012)
doi:10.1103/PhysRevD.85.104048
[arXiv:1202.5800 [gr-qc]].
%29 citations counted in INSPIRE as of 04 Jun 2023

%\cite{Schwinger:1960qe}
\bibitem{Schwinger:1960qe}
J.~S.~Schwinger,
%``Brownian motion of a quantum oscillator,''
J. Math. Phys. \textbf{2}, 407-432 (1961)
doi:10.1063/1.1703727
%1758 citations counted in INSPIRE as of 04 Jun 2023

%\cite{Mahanthappa:1962ex}
\bibitem{Mahanthappa:1962ex}
K.~T.~Mahanthappa,
%``Multiple production of photons in quantum electrodynamics,''
Phys. Rev. \textbf{126}, 329-340 (1962)
doi:10.1103/PhysRev.126.329
%319 citations counted in INSPIRE as of 04 Jun 2023

%\cite{Bakshi:1962dv}
\bibitem{Bakshi:1962dv}
P.~M.~Bakshi and K.~T.~Mahanthappa,
%``Expectation value formalism in quantum field theory. 1.,''
J. Math. Phys. \textbf{4}, 1-11 (1963)
doi:10.1063/1.1703883
%371 citations counted in INSPIRE as of 04 Jun 2023

%\cite{Bakshi:1963bn}
\bibitem{Bakshi:1963bn}
P.~M.~Bakshi and K.~T.~Mahanthappa,
%``Expectation value formalism in quantum field theory. 2.,''
J. Math. Phys. \textbf{4}, 12-16 (1963)
doi:10.1063/1.1703879
%333 citations counted in INSPIRE as of 04 Jun 2023

%\cite{Keldysh:1964ud}
\bibitem{Keldysh:1964ud}
L.~V.~Keldysh,
%``Diagram technique for nonequilibrium processes,''
Zh. Eksp. Teor. Fiz. \textbf{47}, 1515-1527 (1964)
%1812 citations counted in INSPIRE as of 04 Jun 2023

%\cite{Chou:1984es}
\bibitem{Chou:1984es}
K.~c.~Chou, Z.~b.~Su, B.~l.~Hao and L.~Yu,
%``Equilibrium and Nonequilibrium Formalisms Made Unified,''
Phys. Rept. \textbf{118}, 1-131 (1985)
doi:10.1016/0370-1573(85)90136-X
%871 citations counted in INSPIRE as of 04 Jun 2023

%\cite{Jordan:1986ug}
\bibitem{Jordan:1986ug}
R.~D.~Jordan,
%``Effective Field Equations for Expectation Values,''
Phys. Rev. D \textbf{33}, 444-454 (1986)
doi:10.1103/PhysRevD.33.444
%447 citations counted in INSPIRE as of 04 Jun 2023

%\cite{Calzetta:1986ey}
\bibitem{Calzetta:1986ey}
E.~Calzetta and B.~L.~Hu,
%``Closed Time Path Functional Formalism in Curved Space-Time: Application to Cosmological Back Reaction Problems,''
Phys. Rev. D \textbf{35}, 495 (1987)
doi:10.1103/PhysRevD.35.495
%488 citations counted in INSPIRE as of 04 Jun 2023

%\cite{Ford:2004wc}
\bibitem{Ford:2004wc}
L.~H.~Ford and R.~P.~Woodard,
%``Stress tensor correlators in the Schwinger-Keldysh formalism,''
Class. Quant. Grav. \textbf{22}, 1637-1647 (2005)
doi:10.1088/0264-9381/22/9/011
[arXiv:gr-qc/0411003 [gr-qc]].
%59 citations counted in INSPIRE as of 04 Jun 2023

%\cite{Miao:2017feh}
\bibitem{Miao:2017feh}
S.~P.~Miao, T.~Prokopec and R.~P.~Woodard,
%``Deducing Cosmological Observables from the S-matrix,''
Phys. Rev. D \textbf{96}, no.10, 104029 (2017)
doi:10.1103/PhysRevD.96.104029
[arXiv:1708.06239 [gr-qc]].
%17 citations counted in INSPIRE as of 01 Jun 2023

%\cite{Katuwal:2020rkv}
\bibitem{Katuwal:2020rkv}
S.~Katuwal and R.~P.~Woodard,
%``Gauge independent quantum gravitational corrections to Maxwell\textquoteright{}s equation,''
JHEP \textbf{21}, 029 (2020)
doi:10.1007/JHEP10(2021)029
[arXiv:2107.13341 [gr-qc]].
%3 citations counted in INSPIRE as of 01 Jun 2023

%\cite{Weinberg:1965nx}
\bibitem{Weinberg:1965nx}
S.~Weinberg,
%``Infrared photons and gravitons,''
Phys. Rev. \textbf{140}, B516-B524 (1965)
doi:10.1103/PhysRev.140.B516
%1006 citations counted in INSPIRE as of 05 Jun 2023

%\cite{Veneziano:1972rs}
\bibitem{Veneziano:1972rs}
G.~Veneziano,
%``Trilinear coupling of scalar bosons in the small mass limit,''
Nucl. Phys. B \textbf{44}, 142-148 (1972)
doi:10.1016/0550-3213(72)90275-1
%22 citations counted in INSPIRE as of 05 Jun 2023

%\cite{Marolf:2012kh}
\bibitem{Marolf:2012kh}
D.~Marolf, I.~A.~Morrison and M.~Srednicki,
%``Perturbative S-matrix for massive scalar fields in global de Sitter space,''
Class. Quant. Grav. \textbf{30}, 155023 (2013)
doi:10.1088/0264-9381/30/15/155023
[arXiv:1209.6039 [hep-th]].
%65 citations counted in INSPIRE as of 05 Jun 2023

%\cite{Miao:2021gic}
\bibitem{Miao:2021gic}
S.~P.~Miao, N.~C.~Tsamis and R.~P.~Woodard,
%``Summing inflationary logarithms in nonlinear sigma models,''
JHEP \textbf{03}, 069 (2022)
doi:10.1007/JHEP03(2022)069
[arXiv:2110.08715 [gr-qc]].
%2 citations counted in INSPIRE as of 24 Apr 2022

%\cite{Borchers:1960}
\bibitem{Borchers:1960}
H.~J.~Borchers,
%``Uber die Mannigfaltigkeit der interpolierenden Felder zu einer kausalen S-Matrix,''
Il Nuovo Cimento \textbf{15}, 784-794 (1960)
doi.org/10.1007/BF02732693.

%\cite{Woodard:2023rqo}
\bibitem{Woodard:2023rqo}
R.~P.~Woodard and B.~Yesilyurt,
%``Unfinished Business in A Nonlinear Sigma Model on de Sitter Background,''
[arXiv:2302.11528 [gr-qc]].
%1 citations counted in INSPIRE as of 24 May 2023

%\cite{Glavan:2013jca}
\bibitem{Glavan:2013jca}
D.~Glavan, S.~P.~Miao, T.~Prokopec and R.~P.~Woodard,
%``Electrodynamic Effects of Inflationary Gravitons,''
Class. Quant. Grav. \textbf{31}, 175002 (2014)
doi:10.1088/0264-9381/31/17/175002
[arXiv:1308.3453 [gr-qc]].
%36 citations counted in INSPIRE as of 05 Jun 2023

%\cite{Wang:2014tza}
\bibitem{Wang:2014tza}
C.~L.~Wang and R.~P.~Woodard,
%``Excitation of Photons by Inflationary Gravitons,''
Phys. Rev. D \textbf{91}, no.12, 124054 (2015)
doi:10.1103/PhysRevD.91.124054
[arXiv:1408.1448 [gr-qc]].
%30 citations counted in INSPIRE as of 05 Jun 2023

%\cite{Miao:2006gj}
\bibitem{Miao:2006gj}
S.~P.~Miao and R.~P.~Woodard,
%``Gravitons Enhance Fermions during Inflation,''
Phys. Rev. D \textbf{74}, 024021 (2006)
doi:10.1103/PhysRevD.74.024021
[arXiv:gr-qc/0603135 [gr-qc]].
%79 citations counted in INSPIRE as of 05 Jun 2023

%\cite{Glavan:2021adm}
\bibitem{Glavan:2021adm}
D.~Glavan, S.~P.~Miao, T.~Prokopec and R.~P.~Woodard,
%``Large logarithms from quantum gravitational corrections to a massless, minimally coupled scalar on de Sitter,''
JHEP \textbf{03}, 088 (2022)
doi:10.1007/JHEP03(2022)088
[arXiv:2112.00959 [gr-qc]].
%8 citations counted in INSPIRE as of 05 Jun 2023

%\cite{Tan:2022xpn}
\bibitem{Tan:2022xpn}
L.~Tan, N.~C.~Tsamis and R.~P.~Woodard,
%``How Inflationary Gravitons Affect the Force of Gravity,''
Universe \textbf{8}, no.7, 376 (2022)
doi:10.3390/universe8070376
[arXiv:2206.11467 [gr-qc]].
%4 citations counted in INSPIRE as of 05 Jun 2023

%\cite{Onemli:2002hr}
\bibitem{Onemli:2002hr}
V.~K.~Onemli and R.~P.~Woodard,
%``Superacceleration from massless, minimally coupled phi**4,''
Class. Quant. Grav. \textbf{19}, 4607 (2002)
doi:10.1088/0264-9381/19/17/311
[arXiv:gr-qc/0204065 [gr-qc]].
%390 citations counted in INSPIRE as of 05 Jun 2023

%\cite{Onemli:2004mb}
\bibitem{Onemli:2004mb}
V.~K.~Onemli and R.~P.~Woodard,
%``Quantum effects can render w \ensuremath{<} -1 on cosmological scales,''
Phys. Rev. D \textbf{70}, 107301 (2004)
doi:10.1103/PhysRevD.70.107301
[arXiv:gr-qc/0406098 [gr-qc]].
%499 citations counted in INSPIRE as of 05 Jun 2023

%\cite{Starobinsky:1986fx}
\bibitem{Starobinsky:1986fx}
A.~A.~Starobinsky,
%``STOCHASTIC DE SITTER (INFLATIONARY) STAGE IN THE EARLY UNIVERSE,''
Lect. Notes Phys. \textbf{246}, 107-126 (1986)
doi:10.1007/3-540-16452-9\_6
%327 citations counted in INSPIRE as of 05 Jun 2023

%\cite{Starobinsky:1994bd}
\bibitem{Starobinsky:1994bd}
A.~A.~Starobinsky and J.~Yokoyama,
%``Equilibrium state of a selfinteracting scalar field in the De Sitter background,''
Phys. Rev. D \textbf{50}, 6357-6368 (1994)
doi:10.1103/PhysRevD.50.6357
[arXiv:astro-ph/9407016 [astro-ph]].
%648 citations counted in INSPIRE as of 05 Jun 2023

%\cite{Litos:2023}
\bibitem{Litos:2023}
C.~Litos, R.~P.~Woodard and B.~Yesilyurt,
%``Large Inflationary Logarithms in a Nontrivial Nonlinear Sigma Model,''
University of Florida preprint UFIFT-QG-23-07, 
in preparation.

%\cite{Pauli:1956}
\bibitem{Pauli:1956}
W.~Pauli, 
%``Comment to Generalization of Einstein's Theory of Gravitation Considered from the Point of View of Quantum Field Theory by O. Klein,''
Helv. Phys. Acta Suppl. {\bf 4}, 58-71 (1956)

%\cite{Deser:1957zz}
\bibitem{Deser:1957zz}
S.~Deser,
%``General Relativity and the Divergence Problem in Quantum Field Theory,''
Rev. Mod. Phys. \textbf{29}, 417 (1957)
doi:10.1103/RevModPhys.29.417
%108 citations counted in INSPIRE as of 06 Jun 2023

%\cite{Ford:2005rs}
\bibitem{Ford:2005rs}
L.~H.~Ford,
%``Stochastic spacetime and Brownian motion of test particles,''
Int. J. Theor. Phys. \textbf{44}, 1753-1768 (2005)
doi:10.1007/s10773-005-8893-z
[arXiv:gr-qc/0501081 [gr-qc]].
%25 citations counted in INSPIRE as of 10 Jun 2023

%\cite{Marunovic:2011zw}
\bibitem{Marunovic:2011zw}
A.~Marunovic and T.~Prokopec,
%``Time transients in the quantum corrected Newtonian potential induced by a massless nonminimally coupled scalar field,''
Phys. Rev. D \textbf{83}, 104039 (2011)
doi:10.1103/PhysRevD.83.104039
[arXiv:1101.5059 [gr-qc]].
%20 citations counted in INSPIRE as of 09 Jun 2023

%\cite{Leonard:2013xsa}
\bibitem{Leonard:2013xsa}
K.~E.~Leonard and R.~P.~Woodard,
%``Graviton Corrections to Vacuum Polarization during Inflation,''
Class. Quant. Grav. \textbf{31}, 015010 (2014)
doi:10.1088/0264-9381/31/1/015010
[arXiv:1304.7265 [gr-qc]].
%40 citations counted in INSPIRE as of 05 Jun 2023

%\cite{Radkowski:1970}
\bibitem{Radkowski:1970}
A.~F.~Radkowski,
%``Some aspects of the source description of gravitation,''
Ann. Phys. \textbf{56}, 319-354 (1970)
doi:10.1016/0003-4916(70)90021-7

%\cite{Kasdagli:2023nzj}
\bibitem{Kasdagli:2023nzj}
E.~Kasdagli, M.~Ulloa and R.~P.~Woodard,
%``Coincident massless, minimally coupled scalar correlators on general cosmological backgrounds,''
Phys. Rev. D \textbf{107}, no.10, 105023 (2023)
doi:10.1103/PhysRevD.107.105023
[arXiv:2302.04808 [gr-qc]].
%2 citations counted in INSPIRE as of 10 Jun 2023

%\cite{Woodard:2023cqi}
\bibitem{Woodard:2023cqi}
R.~P.~Woodard and B.~Yesilyurt,
%``Remembrance of Things Past,''
[arXiv:2305.17641 [gr-qc]].
%0 citations counted in INSPIRE as of 10 Jun 2023

%\cite{Horava:2008ih}
\bibitem{Horava:2008ih}
P.~Horava,
%``Membranes at Quantum Criticality,''
JHEP \textbf{03}, 020 (2009)
doi:10.1088/1126-6708/2009/03/020
[arXiv:0812.4287 [hep-th]].
%721 citations counted in INSPIRE as of 06 Jun 2023

%\cite{Horava:2009uw}
\bibitem{Horava:2009uw}
P.~Horava,
%``Quantum Gravity at a Lifshitz Point,''
Phys. Rev. D \textbf{79}, 084008 (2009)
doi:10.1103/PhysRevD.79.084008
[arXiv:0901.3775 [hep-th]].
%2328 citations counted in INSPIRE as of 06 Jun 2023

%\cite{Mukohyama:2010xz}
\bibitem{Mukohyama:2010xz}
S.~Mukohyama,
%``Horava-Lifshitz Cosmology: A Review,''
Class. Quant. Grav. \textbf{27}, 223101 (2010)
doi:10.1088/0264-9381/27/22/223101
[arXiv:1007.5199 [hep-th]].
%283 citations counted in INSPIRE as of 06 Jun 2023

%\cite{Barvinsky:2015kil}
\bibitem{Barvinsky:2015kil}
A.~O.~Barvinsky, D.~Blas, M.~Herrero-Valea, S.~M.~Sibiryakov and C.~F.~Steinwachs,
%``Renormalization of Ho\v{r}ava gravity,''
Phys. Rev. D \textbf{93}, no.6, 064022 (2016)
doi:10.1103/PhysRevD.93.064022
[arXiv:1512.02250 [hep-th]].
%135 citations counted in INSPIRE as of 06 Jun 2023

%\cite{Carlip:2012wa}
\bibitem{Carlip:2012wa}
S.~Carlip,
%``Challenges for Emergent Gravity,''
Stud. Hist. Phil. Sci. B \textbf{46}, 200-208 (2014)
doi:10.1016/j.shpsb.2012.11.002
[arXiv:1207.2504 [gr-qc]].
%64 citations counted in INSPIRE as of 06 Jun 2023

%\cite{Verlinde:2016toy}
\bibitem{Verlinde:2016toy}
E.~P.~Verlinde,
%``Emergent Gravity and the Dark Universe,''
SciPost Phys. \textbf{2}, no.3, 016 (2017)
doi:10.21468/SciPostPhys.2.3.016
[arXiv:1611.02269 [hep-th]].
%349 citations counted in INSPIRE as of 06 Jun 2023

%\cite{Hossenfelder:2017eoh}
\bibitem{Hossenfelder:2017eoh}
S.~Hossenfelder,
%``Covariant version of Verlinde\textquoteright{}s emergent gravity,''
Phys. Rev. D \textbf{95}, no.12, 124018 (2017)
doi:10.1103/PhysRevD.95.124018
[arXiv:1703.01415 [gr-qc]].
%55 citations counted in INSPIRE as of 06 Jun 2023

%\cite{Ambjorn:2012jv}
\bibitem{Ambjorn:2012jv}
J.~Ambjorn, A.~Goerlich, J.~Jurkiewicz and R.~Loll,
%``Nonperturbative Quantum Gravity,''
Phys. Rept. \textbf{519}, 127-210 (2012)
doi:10.1016/j.physrep.2012.03.007
[arXiv:1203.3591 [hep-th]].
%376 citations counted in INSPIRE as of 06 Jun 2023

%\cite{Ambjorn:2020rcn}
\bibitem{Ambjorn:2020rcn}
J.~Ambjorn, J.~Gizbert-Studnicki, A.~G\"orlich, J.~Jurkiewicz and R.~Loll,
%``Renormalization in quantum theories of geometry,''
Front. in Phys. \textbf{8}, 247 (2020)
doi:10.3389/fphy.2020.00247
[arXiv:2002.01693 [hep-th]].
%25 citations counted in INSPIRE as of 06 Jun 2023

%\cite{Dvali:2010jz}
\bibitem{Dvali:2010jz}
G.~Dvali, G.~F.~Giudice, C.~Gomez and A.~Kehagias,
%``UV-Completion by Classicalization,''
JHEP \textbf{08}, 108 (2011)
doi:10.1007/JHEP08(2011)108
[arXiv:1010.1415 [hep-ph]].
%253 citations counted in INSPIRE as of 06 Jun 2023

%\cite{Dvali:2014ila}
\bibitem{Dvali:2014ila}
G.~Dvali, C.~Gomez, R.~S.~Isermann, D.~L\"ust and S.~Stieberger,
%``Black hole formation and classicalization in ultra-Planckian 2\textrightarrow{}N scattering,''
Nucl. Phys. B \textbf{893}, 187-235 (2015)
doi:10.1016/j.nuclphysb.2015.02.004
[arXiv:1409.7405 [hep-th]].
%109 citations counted in INSPIRE as of 06 Jun 2023

%\cite{Arnowitt:1960zz}
\bibitem{Arnowitt:1960zz}
R.~Arnowitt, S.~Deser and C.~W.~Misner,
%``Finite Self-Energy of Classical Point Particles,''
Phys. Rev. Lett. \textbf{4}, 375-377 (1960)
doi:10.1103/PhysRevLett.4.375
%60 citations counted in INSPIRE as of 24 Apr 2022

%\cite{Woodard:1997dm}
\bibitem{Woodard:1997dm}
R.~P.~Woodard,
%``Particles as bound states in their own potentials,''
[arXiv:gr-qc/9803096 [gr-qc]].
%5 citations counted in INSPIRE as of 30 May 2023

%\cite{Casadio:2009eh}
\bibitem{Casadio:2009eh}
R.~Casadio,
%``Gravitational renormalization of quantum field theory: A 'Conservative' approach,''
J. Phys. Conf. Ser. \textbf{174}, 012058 (2009)
doi:10.1088/1742-6596/174/1/012058
[arXiv:0902.2939 [gr-qc]].
%2 citations counted in INSPIRE as of 30 May 2023

%\cite{Casadio:2009jc}
\bibitem{Casadio:2009jc}
R.~Casadio, R.~Garattini and F.~Scardigli,
%``Point-like sources and the scale of quantum gravity,''
Phys. Lett. B \textbf{679}, 156-159 (2009)
doi:10.1016/j.physletb.2009.06.076
[arXiv:0904.3406 [gr-qc]].
%26 citations counted in INSPIRE as of 30 May 2023

%\cite{Mora:2011jj}
\bibitem{Mora:2011jj}
P.~J.~Mora, N.~C.~Tsamis and R.~P.~Woodard,
%``Generalizing the ADM Computation to Quantum Field Theory,''
Class. Quant. Grav. \textbf{29}, 025001 (2012)
doi:10.1088/0264-9381/29/2/025001
[arXiv:1108.4367 [gr-qc]].
%4 citations counted in INSPIRE as of 30 May 2023

%\cite{Arnowitt:1959ah}
\bibitem{Arnowitt:1959ah}
R.~L.~Arnowitt, S.~Deser and C.~W.~Misner,
%``Dynamical Structure and Definition of Energy in General Relativity,''
Phys. Rev. \textbf{116}, 1322-1330 (1959)
doi:10.1103/PhysRev.116.1322
%712 citations counted in INSPIRE as of 08 Jun 2023

%\cite{Maldacena:2002vr}
\bibitem{Maldacena:2002vr}
J.~M.~Maldacena,
%``Non-Gaussian features of primordial fluctuations in single field inflationary models,''
JHEP \textbf{05}, 013 (2003)
doi:10.1088/1126-6708/2003/05/013
[arXiv:astro-ph/0210603 [astro-ph]].
%2610 citations counted in INSPIRE as of 08 Jun 2023

%\cite{Seery:2006vu}
\bibitem{Seery:2006vu}
D.~Seery, J.~E.~Lidsey and M.~S.~Sloth,
%``The inflationary trispectrum,''
JCAP \textbf{01}, 027 (2007)
doi:10.1088/1475-7516/2007/01/027
[arXiv:astro-ph/0610210 [astro-ph]].
%195 citations counted in INSPIRE as of 08 Jun 2023

%\cite{Jarnhus:2007ia}
\bibitem{Jarnhus:2007ia}
P.~R.~Jarnhus and M.~S.~Sloth,
%``de Sitter limit of inflation and nonlinear perturbation theory,''
JCAP \textbf{02}, 013 (2008)
doi:10.1088/1475-7516/2008/02/013
[arXiv:0709.2708 [hep-th]].
%40 citations counted in INSPIRE as of 08 Jun 2023

%\cite{Xue:2012wi}
\bibitem{Xue:2012wi}
W.~Xue, X.~Gao and R.~Brandenberger,
%``IR Divergences in Inflation and Entropy Perturbations,''
JCAP \textbf{06}, 035 (2012)
doi:10.1088/1475-7516/2012/06/035
[arXiv:1201.0768 [hep-th]].
%41 citations counted in INSPIRE as of 08 Jun 2023

%\cite{Starobinsky:2007hu}
\bibitem{Starobinsky:2007hu}
A.~A.~Starobinsky,
%``Disappearing cosmological constant in f(R) gravity,''
JETP Lett. \textbf{86}, 157-163 (2007)
doi:10.1134/S0021364007150027
[arXiv:0706.2041 [astro-ph]].
%1085 citations counted in INSPIRE as of 24 May 2023

\end{thebibliography}
\end{document}